\documentclass[aps,prd,onecolumn,groupedaddress,nofootinbib,superscriptaddress]{revtex4}  
\usepackage{graphicx,mathtools,tabu}
\usepackage{epstopdf}
\usepackage{amsmath}
\usepackage{amsfonts}
\usepackage[colorlinks=true,citecolor=red,linkcolor=blue,breaklinks=true]{hyperref}
\usepackage[figure, table]{hypcap}
\usepackage{amssymb}
\usepackage{appendix}
\usepackage{comment}
\usepackage{bbold}
\usepackage{color}
\usepackage{slashed}
\usepackage{subfigure}
\usepackage{setspace}
\usepackage{footnote}
\usepackage{multirow}
\usepackage{braket}
\usepackage[normalem]{ulem}
\usepackage[utf8]{inputenc}
\usepackage{mathrsfs}
\usepackage{longtable}
\usepackage{enumitem}

\LTcapwidth=\textwidth

\usepackage{url}
\usepackage{wrapfig}
\usepackage{braket}

 \def\be   {\begin{equation}}  
  \def\ee   {\end{equation}}
 \def\ba   {\begin{array}}     
  \def\ea   {\end{array}}
 \def\bea  {\begin{eqnarray}}  
  \def\eea  {\end{eqnarray}}
 \def\bean {\begin{eqnarray*}}  
 \def\eean {\end{eqnarray*}}
 \def\ga   {\gamma}

  \def\al   {\alpha}
  \def\be {\beta}
    
      \def\r {\rho}
     
  \def\sig   {\sigma}

\def\to {\rightarrow}
\def\Lam{\Lambda}

\definecolor{darkgreen}{rgb}{0,0.5,0}

\begin{document}

\hfill {NUHEP-TH/19-07, CP3-Origins-2019-027 DNRF90}

\title{Accessible Lepton-Number-Violating Models and Negligible Neutrino Masses}

\author{Andr\'{e} de Gouv\^{e}a}
\email{degouvea@northwestern.edu}
\affiliation{Northwestern University, Department of Physics \& Astronomy, 2145 Sheridan Road, Evanston, IL 60208, USA}
\author{Wei-Chih Huang}
\email{huang@cp3.sdu.dk}
\affiliation{CP$^{3}$-Origins, University of Southern Denmark, Campusvej 55, DK-5230 Odense M, Denmark}
\author{Johannes K\"onig}
\email{konig@cp3.sdu.dk}
\affiliation{CP$^{3}$-Origins, University of Southern Denmark, Campusvej 55, DK-5230 Odense M, Denmark}
\author{Manibrata Sen}
\email{manibrata@berkeley.edu}
\affiliation{Northwestern University, Department of Physics \& Astronomy, 2145 Sheridan Road, Evanston, IL 60208, USA}
\affiliation{Department of Physics, University of California Berkeley, Berkeley, California 94720, USA}

\begin{abstract}
Lepton-number violation (LNV), in general, implies nonzero Majorana masses for the Standard Model neutrinos. Since neutrino masses are very small, for generic candidate models of the physics responsible for LNV, the rates for almost all  experimentally accessible LNV observables -- except for neutrinoless double-beta decay -- are expected to be exceedingly small. Guided by effective-operator considerations of LNV phenomena, we identify a complete family of models where lepton number is violated but the generated Majorana neutrino masses are tiny, even if the new-physics scale is below 1~TeV. We explore the phenomenology of these models, including charged-lepton flavor-violating phenomena and baryon-number-violating phenomena, identifying scenarios where the allowed rates for $\mu^-\to e^+$-conversion in nuclei are potentially accessible to next-generation experiments. 
\end{abstract}

\maketitle
\section{Introduction}
\label{sec:intro}

Lepton number and baryon number are, at the classical level, accidental global symmetries of the renormalizable Standard Model (SM) Lagrangian.\footnote{At the quantum level these symmetries are anomalous, i.e., they are violated by non-perturbative effects \cite{tHooft:1976rip,Klinkhamer:1984di}. These are only relevant in extraordinary circumstances (e.g., very high temperatures) much beyond the reach of particle physics experiments \cite{Ellis:2016ast,Tye:2015tva}. Non-perturbative effects still preserve baryon-number-minus-lepton number, the non-anomalous possible global symmetry of the renormalizable SM Lagrangian.} If one allows for generic non-renormalizable operators consistent with the SM gauge symmetries and particle content, lepton number and baryon number will no longer be conserved. Indeed, lepton-number conservation is violated by effective operators of dimension five or higher while baryon-number conservation (sometimes together with lepton number) is violated by effective operators of dimension six or higher. In other words, generically, the addition of new degrees-of-freedom to the SM particle content violates baryon-number and lepton-number conservation.

Experimentally, in spite of ambitious ongoing experimental efforts, there is no evidence for the violation of lepton-number or baryon-number conservation \cite{Tanabashi:2018oca}. There are a few different potential explanations for these (negative) experimental results, assuming degrees-of-freedom beyond those of the SM exist. Perhaps the new particles are either very heavy or very weakly coupled in such a way that phenomena that violate lepton-number or baryon-number conservation are highly suppressed. Another possibility is that the new interactions are not generic and that lepton-number or baryon-number conservation are global symmetries of the Beyond-the-Standard-Model Lagrangian. Finally, it is possible that even though baryon number or lepton number are not conserved and the new degrees-of-freedom are neither weakly coupled nor very heavy, only a subset of baryon-number-violating or lepton-number-violating phenomena are within reach of particle physics experiments. This manuscript concentrates on this third option, which we hope to elucidate below. 

The discovery of nonzero yet tiny neutrino masses is often interpreted as enticing -- but certainly not definitive! -- indirect evidence for lepton-number-violating new physics. In this case, neutrinos are massive Majorana fermions and one can naturally ``explain'' why the masses of neutrinos are much smaller than those of all other known massive particles (see, for example, \cite{Schechter:1981cv,deGouvea:2013zba,Gouvea:2016shl} for discussions of this point). Searches for the nature of the neutrino -- Majorana fermion versus Dirac fermion -- are most often searches for lepton number violation (LNV). The observation of LNV implies, generically, that neutrinos are Majorana fermions \cite{Schechter:1981bd}, while Majorana neutrino masses imply nonzero rates for lepton-number-violating phenomena. The most powerful probes of LNV are searches for neutrinoless double-beta decay ($0\nu\beta\beta$, see \cite{Rodejohann:2011mu} for a review); several of these are ongoing, for example  \cite{Albert:2014awa,Gando:2012zm,GERDA:2018zzh}. The growing excitement behind searches for $0\nu\beta\beta$ is the fact that these are sensitive enough to detect LNV mediated by light Majorana neutrino exchange if the neutrino masses are above a fraction on an electronvolt. In many models that lead to Majorana neutrino masses, including, arguably, the simplest, most elegant, and best motivated ones, LNV phenomena are predominantly mediated by light Majorana neutrino exchange. Hence, we are approaching sensitivities to $0\nu\beta\beta$ capable of providing nontrivial, robust information on the nature of the neutrino. 

Other searches for LNV are, in general, not as sensitive as those for $0\nu\beta\beta$. Here, we will highlight searches for $\mu^-\to e^+$-conversion in nuclei, for a couple of reasons. One is that, except for searches for $0\nu\beta\beta$, searches for $\mu^-\to e^+$-conversion in nuclei are, arguably, the most sensitive to generic LNV new physics.\footnote{It was recently pointed out that searches for non-standard neutrino interactions from long-baseline neutrino experiments are also sensitive to certain LNV new physics and involve all lepton flavors~\cite{Bolton:2019wta}. In some cases, the resulting limits are stronger than those from $\mu^- \to e^+$-conversion.}
Second, several different experiments aimed at searching for $\mu^-\to e^-$-conversion in nuclei are under construction, including the COMET \cite{Kuno:2013mha} and DeeMe experiments \cite{Natori:2014yba} in J-PARC, and the {\it Mu2e} experiment \cite{Bartoszek:2014mya} in Fermilab. These efforts are expected to increase the sensitivity to 
$\mu^-\to e^-$-conversion by, ultimately, four orders of magnitude and may also be able to extend the sensitivity to $\mu^-\to e^+$-conversion in nuclei by at least a few orders of magnitude. 

The best bounds on the $\mu^- \to e^+$-conversion rate relative to the capture rate of a $\mu^-$ on titanium were obtained by the SINDRUM~II experiment \cite{Kaulard:1998rb} over twenty years ago:
\begin{equation}
\label{mutoe}
R_{\mu^- e^+}^\text{Ti} \equiv \frac{\Gamma(\mu^- + \text{Ti} \to e^+ + \text{Ca})}{\Gamma(\mu^- + \text{Ti} \to \nu_\mu + \text{Sc})} < \left\{
\begin{array}{l}
1.7 \times 10^{-12} \text{ (GS, 90\% CL)} \\ 
3.6 \times 10^{-11} \text{ (GDR, 90\% CL)}  \end{array} \right. ,
\end{equation}
where GS considers scattering off titanium to the ground state of calcium, whereas GDR considers the transition to a giant dipole resonance state. Next-generation experiments like {\it Mu2e}, DeeMe, and COMET have the potential to be much more sensitive to $\mu^-\to e^+$-conversion. The authors of \cite{Berryman:2016slh} naively estimated the future sensitivities of these experiments to be
\begin{align}
\text{Mu2e:} & \quad R^{\text{Al}}_{\mu^-e^+} \gtrsim 10^{-16}, \label{mu2ebound}\\
\text{COMET Phase-I:} & \quad R^{\text{Al}}_{\mu^-e^+} \gtrsim 10^{-14}. 
\end{align}
For a recent, more detailed discussion, see \cite{Yeo:2017fej}. 

There are several recent phenomenological attempts at understanding whether there are models consistent with current experimental constraints where the rate for $\mu^-\to e^+$-conversion in nuclei is sizable \cite{Geib:2016atx,Berryman:2016slh,Geib:2016daa}. The main challenges are two-fold. On the one hand, the light-Majorana-neutrino exchange contribution to $\mu^-\to e^+$-conversion in nuclei is tiny. On the other hand, while it is possible to consider other LNV effects that are not captured by light-Majorana-neutrino exchange, most of these scenarios lead, once the new degrees of freedom are integrated out, to Majorana neutrino masses that are way too large and safely excluded by existing neutrino data. In \cite{Berryman:2016slh}, an effective operator approach, introduced and exploited in, for example,~\cite{Babu:2001ex,deGouvea:2007qla,Angel:2012ug,Deppisch:2017ecm}, was employed to both diagnose the problem and identify potentially interesting directions for model building.

New-physics scenarios that violate lepton-number conservation at the tree level in a way that LNV low-energy phenomena are captured by the `all-singlets' dimension-nine operator:
\begin{equation}
{\cal L}  \supset \frac{1}{\Lambda^5}{\mathcal O}_s\,,~~~{\rm where}~~~ {\mathcal O}_s= e^c \mu^c u^c u^c \overline{d^c}\, \overline{d^c}\,,
	\label{eq:effectiveOperator}
\end{equation}
and $\Lambda$ is the effective scale of the operator, were flagged as very ``inefficient'' when it comes to generating neutrino Majorana masses. According to \cite{Berryman:2016slh}, the contribution to Majorana neutrino masses from the physics that leads to Eq.~(\ref{eq:effectiveOperator}) at the tree level saturates the upper bound on neutrino masses for $\Lambda\sim 1$~GeV. This means that, for $\Lambda\gg 1$~GeV, the physics responsible for Eq.~(\ref{eq:effectiveOperator}) will lead to neutrino masses that are too small to be significant while the rates of other LNV phenomena, including $\mu^-\to e^+$-conversion in nuclei, may be within reach of next-generation experiments. According to  \cite{Berryman:2016slh}, this happens for $\Lambda\lesssim 100$~GeV.

The effective operator approach from
\cite{Babu:2001ex,deGouvea:2007qla,Angel:2012ug,Deppisch:2017ecm} is mostly powerless when it
comes to addressing lepton-number-conserving, low-energy effects of the same
physics that leads to Eq.~(\ref{eq:effectiveOperator}). One way to understand
this is to appreciate that lepton-number-conserving phenomena are captured by
qualitatively different effective operators and, in general, it is not possible
to relate different ``types'' of operators in a model-independent way. Concrete
results can only be obtained for ultraviolet (UV)-complete scenarios.

In this manuscript, we systematically identify all possible UV-complete models that are predominantly captured, when it comes to LNV phenomena, by $\mathcal{O}_s$ at the tree level. All these models are expected to have one thing in common: potentially large contributions to LNV processes combined with insignificant contributions to the light neutrino masses. Such models are expected to manifest themselves most efficiently in LNV phenomena like $\mu^-\to e^+$-conversion in nuclei, lepton-number-conserving phenomena, including charged-lepton flavor-violating (CLFV) observables, or baryon-number-violating phenomena, including neutron--antineutron oscillations. Furthermore, if the rates for $\mu^-\to e^+$-conversion in nuclei are indeed close to being accessible, we find that all tree-level realizations of $\mathcal{O}_s$ require the existence of new degrees-of-freedom with masses that are within reach of TeV-scale colliders like the LHC. 

The following sections are organized as follows. In Sec.\,\ref{sec:effOp}, we study the all-singlets effective operator and illustrate its contributions to neutrino masses, $0\nu\beta\beta$, and $\mu^-\to e^+$-conversion in nuclei. In Sec.\,\ref{sec:UV}, we list the different UV-complete models that are associated to the all-singlets effective operator at tree level. We discuss various bounds arising from searches for baryon-number-violating and CLFV processes. In Sec.\,\ref{sec:coll}, we comment on some salient collider signatures of the different new particles proposed in this work. Finally, in Sec.\,\ref{sec:conclusion}, we briefly comment on possible extensions of these scenarios which can account for the observed neutrino masses, summarize our results, and conclude.

\setcounter{equation}{0}
\section{The effective all-singlets operator $O^{\al\be}_s$}
\label{sec:effOp}

In the context of particle physics phenomenology, different notations are prevalent in the literature. Before proceeding, we outline the notation used in this paper, which follows that in \cite{deGouvea:2007qla,Angel:2012ug}. The SM is constructed using only left-chiral Weyl fields: $Q\equiv \bigl(u_L,\,d_L\bigr)$, $L\equiv (\nu_L,\,l_L)$ are the left-chiral ${\rm SU}(2)_L$ doublets, while $u^c,\,d^c$ and $\ell^c$ are the left-chiral ${\rm SU}(2)_L$ singlet fields. The corresponding Hermitian-conjugated fields are identified with a bar (e.g., $\overline{L},\,\overline{e^c}$). Thus, unbarred fields $L_\sigma$ correspond to the $(1/2,0)$ representation of the Lorentz algebra, while barred fields $\overline{L}_{\dot{\sigma}}\equiv L_\sigma^\dagger$ transform under the $(0,1/2)$ representation of the algebra. In this terminology, the familiar four-component Dirac spinor consisting of the electron and the positron can be written as $e = (e_L, \, \overline{e^c} )^T$. Throughout, color indices are implicit and hence omitted. Also, the SM Higgs doublet is taken to be $H\equiv (H^+,\,H^0)^T$, where $H^0$ acquires a vacuum expectation value (vev) $v$ to break the ${\rm SU}(2)_L\times {\rm U}(1)_Y$ gauge-symmetry spontaneously to ${\rm U}(1)_{EM}$.

 Gauge singlets can be formed by either contracting the ${\rm SU}(2)_L$ indices using the antisymmetric tensor $\epsilon_{ij}$ or the Kronecker $\delta_{ij}$ (for conjugated fields). Additionally, flavor couplings are, unless explicitly shown, implicitly contracted. The flavor structure of the effective operators can be used to infer contributions to different new-physics processes, as we shall see. We also do not explicitly show the Lorentz structure of the different operators. Note that, for the same operator, there can be different contractions associated with the gauge and Lorentz indices. These different contractions, however, lead to estimates for the rates of the processes of interest which are roughly the same.
 
 An effective operator of mass dimension $d$ is suppressed by $(d-4)$ powers of the effective mass-scale $\Lambda$ of the new physics, i.e.,
\begin{equation}
 \mathcal{L} \supset \frac{g}{\Lambda^{d-4}}\mathcal{O}^d + h.c.\,,
\end{equation}
where $g\,$s are dimensionless coupling constants. Note that $g$ and $\Lambda$ are not independently defined; one can resolve this issue, e.g., by defining $\Lambda$ such that the largest $g$ is one. The effective scale $\Lambda$ indicates the maximum laboratory energy beyond which the effective-operator description breaks down, i.e., the effective-theory description is valid at energy scales which are at most of order $\Lambda$.

With this arsenal, the dimension-nine all-singlets operators are
\begin{equation}
 \mathcal{O}^{\al\be}_s=\ell_\al^c \ell_\be^c u^c u^c \overline{d^c}\, \overline{d^c}
 \label{eq:effectiveOperatorGen}
\end{equation}
where $\ell^c_\al\equiv\,e^c,\,\mu^c$ or $\tau^c$. $\mathcal{O}_s$ are formed from all the $SU(2)_L$-singlet
fields. If all quarks are of the same generation, there is only one independent Lorentz
contraction: $(\ell_{\alpha}^c)^\sig (\ell_{\beta}^c)_\sig \, (u^c)^\r (u^c)_\r \, (\overline{d^c})_{\dot{\sig}} (\overline{d^c})^{\dot{\sig}}\,$,
where $\sig,\,\rho$ and $\dot{\sig}$ are the Lorentz indices; all other possible contractions are related to this via Fierz transformations.\footnote{If we consider up-type and down-type quarks of different generations, there are two independent contractions. One can choose those to be  $(e^c \mu^c) (u^c t^c) (\overline{d^c}\, \overline{b^c}) \text{ and } (e^c u^c)(\mu^c \sigma^\mu \overline{d^c})(t^c \sigma_\mu \overline{b^c})$, where, for convenience, we fix the two different $\ell^c_{\alpha}$ to be the electron and the muon and the two different generations of quarks to be the first and third generations. All other contractions can be expressed as combinations of these two.}

At different loop-orders, $\mathcal{O}_s^{\al\be}$ will contribute to Majorana neutrino masses as well as different LNV processes. In what follows, we estimate in some detail the contributions of these operators to Majorana neutrino masses, $0\nu\beta\beta$ and $\mu^-\to e^+$-conversion in nuclei. The idea \cite{deGouvea:2007qla} is to start with the effective operator, and add SM interactions to generate the relevant processes. The results presented in this section agree with those in \cite{Berryman:2016slh}.
\begin{figure}[t!]
\centering 
\includegraphics[width=0.5\textwidth]{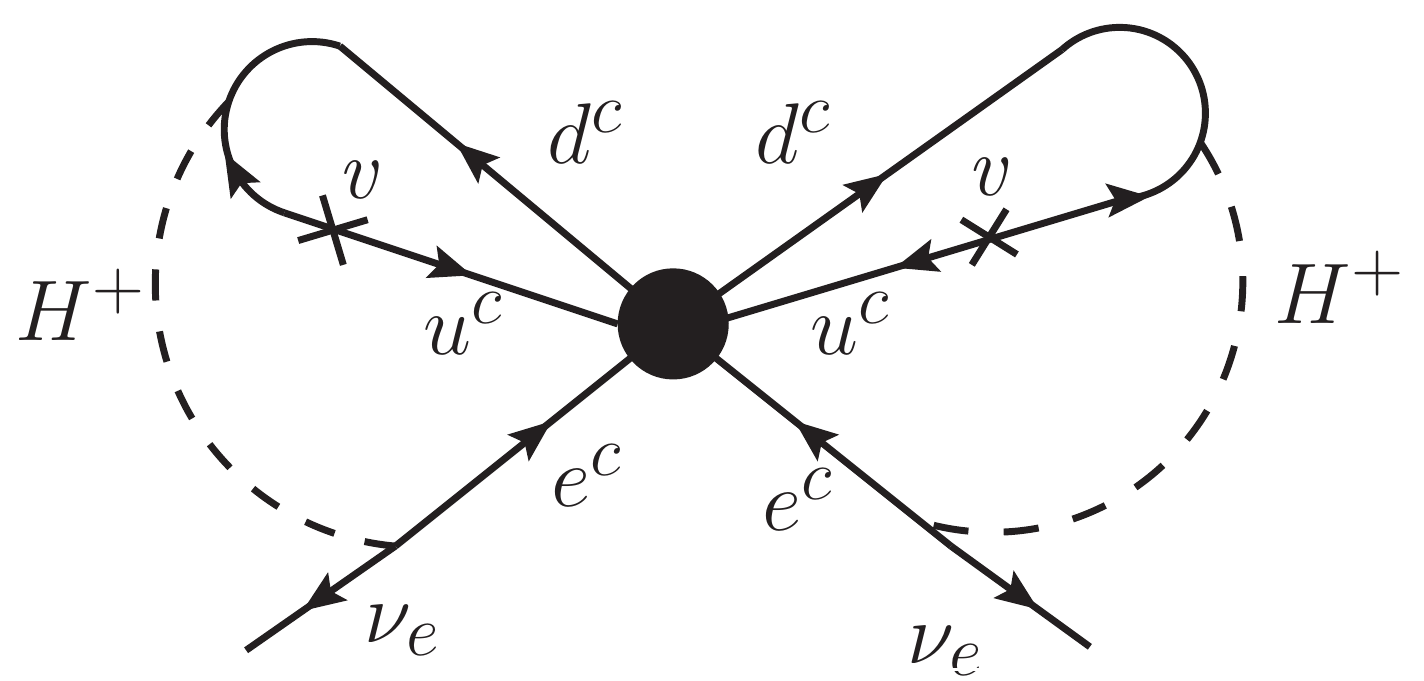}
\caption{Four-loop contribution to Majorana neutrino masses from the dimension-nine all-singlets operator $\mathcal{O}^{ee}_s$. Given the large loop suppression, the contribution to neutrino masses is only relevant for very low effective scales. The blob represents the effective operator while the $\times$ represents the Higgs-boson vacuum expectation value.}
\label{fig:numass}
\end{figure}

Neutrino Majorana masses are generated by the LNV Weinberg operator \cite{Weinberg:1979sa},
\begin{equation}
 \mathcal{L}\supset\frac{f_{\al\be}}{\Lambda_W}\,( L^\al H ) (L^\be H)\,.
\end{equation}
These are dimension-five operators, violate lepton number by two units, and, after electroweak symmetry breaking, lead to neutrino Majorana mass terms, ${\cal L}\supset m_{\al\be}\nu^\al \nu^\be$, $m=fv^2/\Lambda_W$. $\Lambda_W$ is the effective scale of the Weinberg operator, related to but not the same as $\Lambda$, the effective scale of $\mathcal{O}^{\al\be}_s$. Experimental information on neutrino masses point to $\Lambda_{W}\sim 10^{14}$~GeV. Starting from the all-singlets operator,  Fig.~\ref{fig:numass} illustrates how the Weinberg operator is obtained at the four-loop level. In Fig.~\ref{fig:numass}, the blob represents the effective operator $\mathcal{O}^{ee}_s$, for concreteness. Clearly, since $\mathcal{O}^{ee}_s$ involves only $SU(2)_L$-singlet fields, neutrino masses require six Yukawa insertions so one can ``reach'' the corresponding lepton-doublets $L$ and the Higgs-doublet $H$. The contribution to the neutrino mass matrix can be estimated as 
\begin{equation}
 m_{\al\be} = \frac{ g_{ \al \be } }{ \Lambda } \frac{ y_\al y_\be ( y_t y_b v )^2 }{ ( 16 \pi^2 )^4 }\,,
 \label{eq:numass}
\end{equation}
where $y$ are the different charged-lepton and quark Yukawa couplings, $\Lambda$ and $g$ are the effective scale and couplings of $\mathcal{O}^{\al\be}_s$, respectively, and we assumed third-generation quarks, as these are associated to the largest Yukawa couplings. Note that the $\alpha,\beta$ indices in Eq.~(\ref{eq:numass}) are not summed over. 

Neutrino oscillation data constrain only the neutrino mass-squared differences. Nonetheless, one can use the atmospheric and the solar mass-squared differences to set lower bounds on the masses of the heaviest and the next-to-heaviest neutrinos. The atmospheric mass-squared difference, for example, dictates that at least one neutrino has to be heavier than $\sqrt{|\Delta m^2_{32}|}\simeq 0.05 \,{\rm eV}$ \cite{Esteban:2018azc}. On the other hand, cosmic surveys limit the sum of masses of the neutrinos to be $\lesssim 0.12 \,{\rm eV}$ \cite{Vagnozzi:2017ovm, Aghanim:2018eyx,Loureiro:2018pdz}. For concreteness, we assume that the largest element of the neutrino mass matrix lies between $m_\nu \in (0.05-0.5)\,{\rm eV}$. In this case, Eq.\,(\ref{eq:numass}) implies that the effective scale of $\mathcal{O}^{\al\be}_s$ \cite{Berryman:2016slh} is
\begin{equation}
 \Lam\in \left( 100\,{\rm MeV}-1\,{\rm GeV}\right)\,.
 \label{eq:nu_mass}
\end{equation}

\begin{figure}[!t]
\centering 
\includegraphics[width=1\textwidth, height=0.5\textwidth]{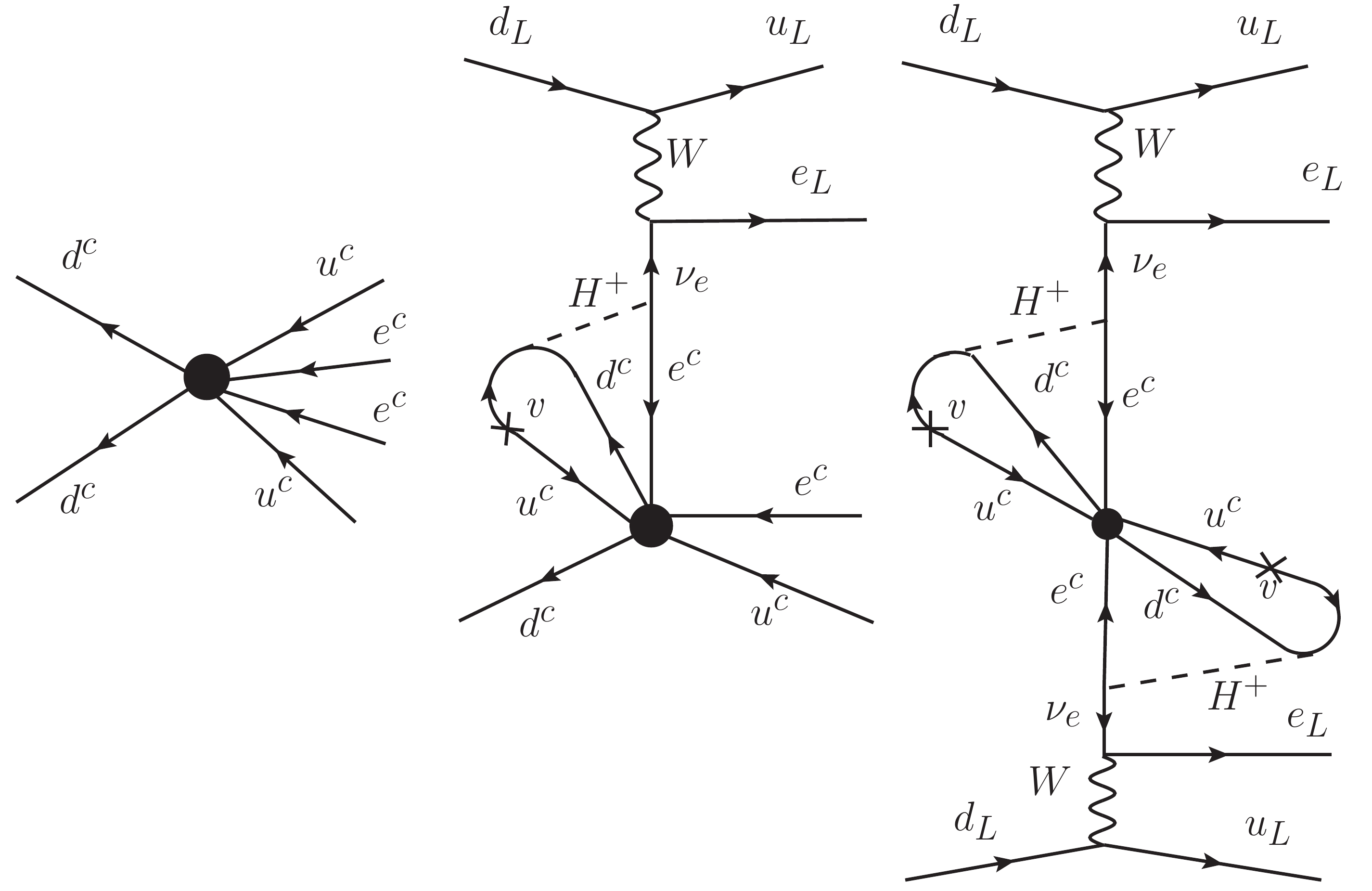}
\caption{Feynman diagrams contributing to $0\nu\beta\beta$ from the dimension-nine all-singlets operator $\mathcal{O}^{ee}_s$ at the tree level (left), two-loop level (middle), and four-loop level (right).  The blob represents the effective operator while the $\times$ represents the Higgs-boson vacuum expectation value.}
\label{fig:NuLessBeta}
\end{figure}
\begin{figure}[!t]
  \centering
  \begin{tabular}{c c}
    \includegraphics[width=0.5\textwidth]{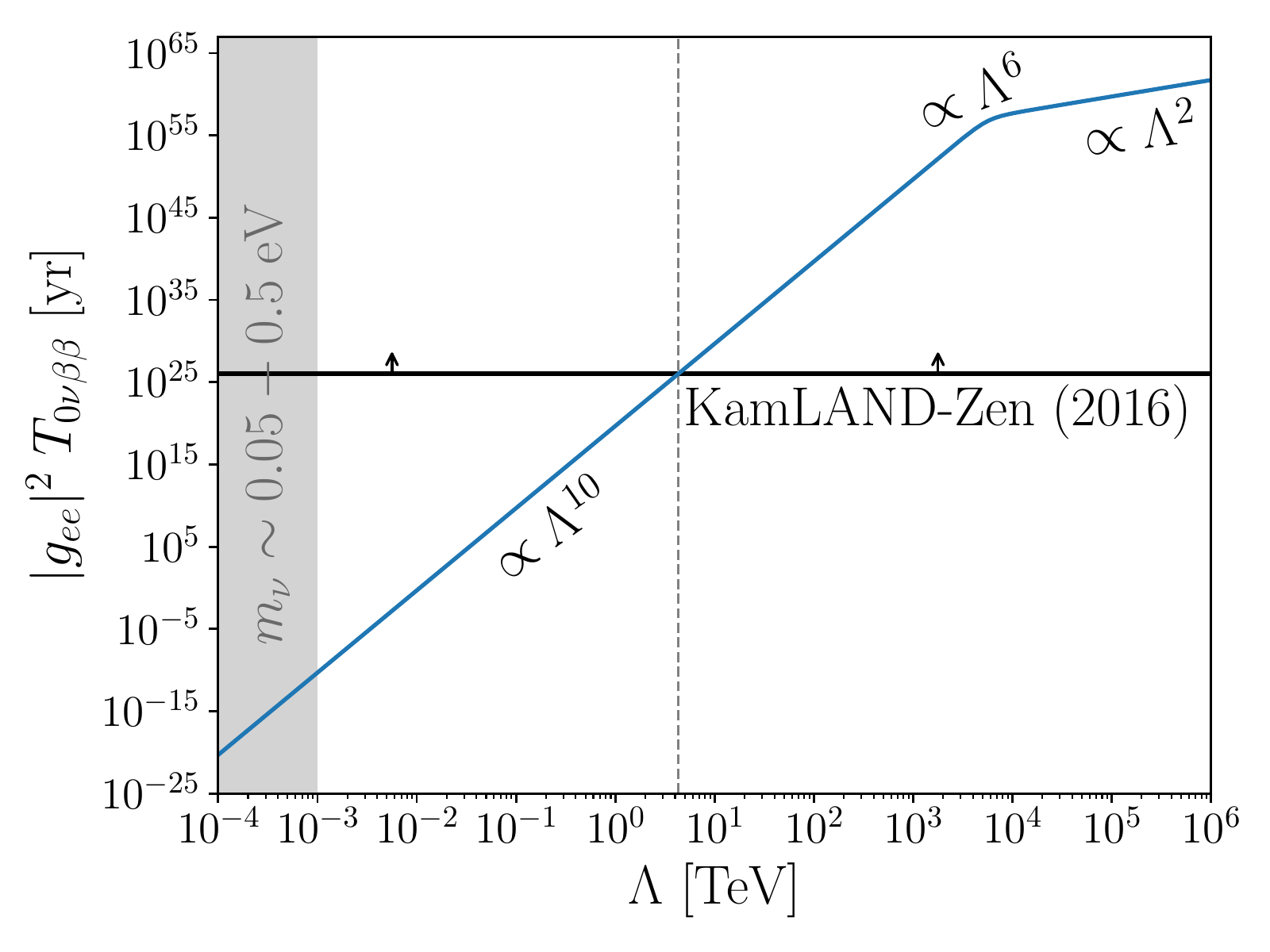}
    \includegraphics[width=0.5\textwidth]{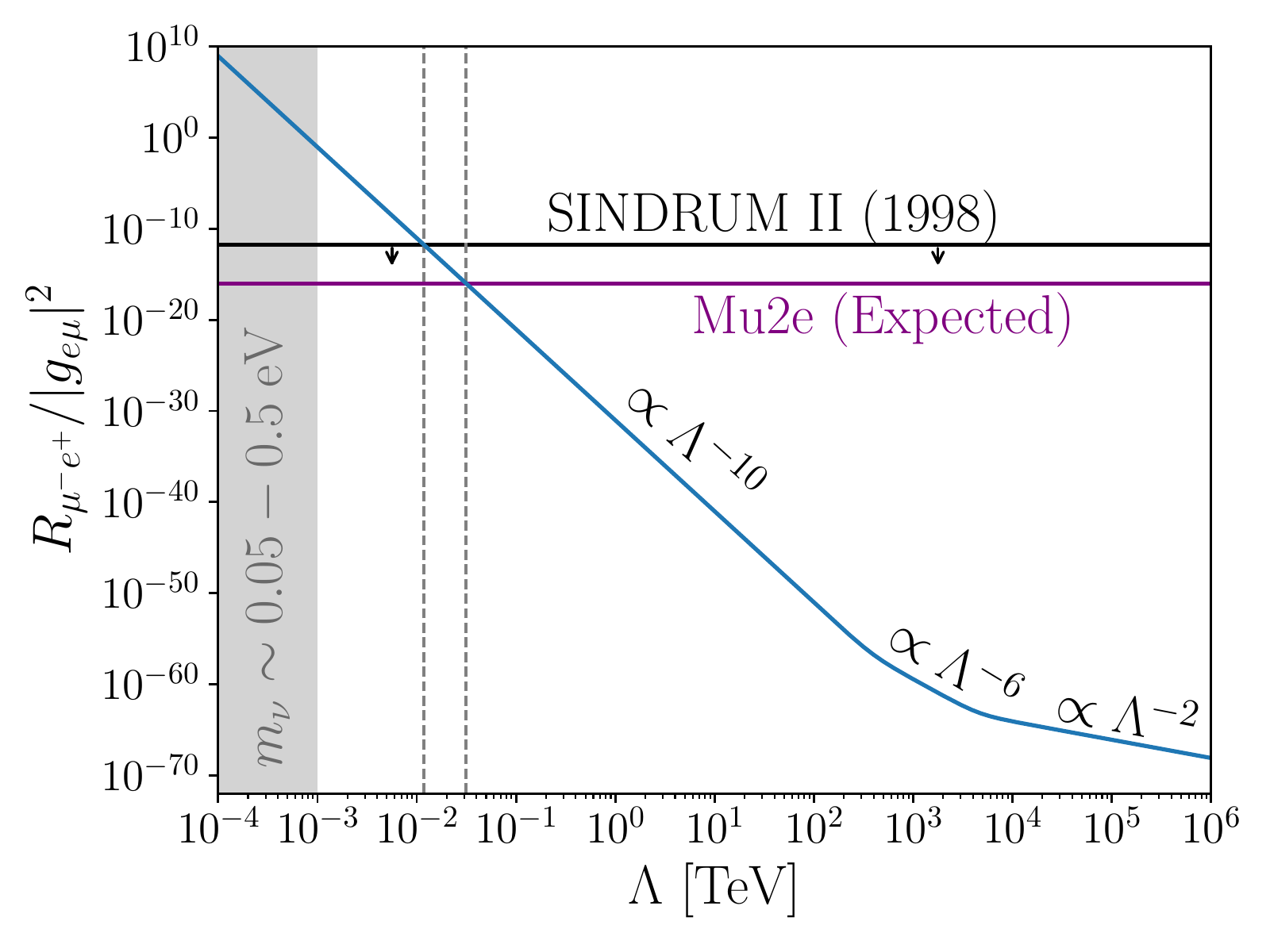}
  \end{tabular}
  \caption{Left: The lifetime associated to $0\nu\beta\beta$,  $T_{0 \nu \be\be}$, as a function of
    the cutoff scale $\Lambda$, from the dimension-nine all-singlets operator $\mathcal{O}^{ee}_s$. For values $\Lambda \lesssim 10^4$~TeV the lifetime
    is dominated by the tree-level contribution and scales like
    $\propto \Lambda^{10}$, whereas for larger values of $\Lambda$, the lifetime
    is dominated by the four-loop contribution and scales $\propto
    \Lambda^2$. The current experimental bound from KamLAND-Zen
    is depicted as a horizontal black line. Right: The normalized rate $R_{\mu^- e^+}$ of muon to positron conversion as
    a function of the cutoff scale $\Lambda$, from the dimension-nine all-singlets operator $\mathcal{O}^{\mu e}_s$. For scales $\Lambda \lesssim
    10^2$~TeV, the tree-level contribution dominates and the the rate scales
    like $\propto \Lambda^{-10}$. For scales $\Lambda \gtrsim 10^4$~TeV the four-loop contribution is most relevant and the rate scales like $\propto
    \Lambda^{-2}$. Between those regions, the two-loop contribuion is most important and the rate scale like $\propto \Lambda^{-6}$. The current experimental bound from SINDRUM II and the sensitivity of {\it Mu2e} are depicted as a horizontal black and purple lines, respectively.}
  \label{fig:Lifetime}
\end{figure}

Fig.~\ref{fig:NuLessBeta} depicts the tree-level, two-loop and four-loop contributions to $0\nu\be\be$ from $\mathcal{O}^{ee}_s$. The half-life for such a decay is estimated as \cite{Berryman:2016slh}
\begin{equation}
T_{0\nu\be\be}=\frac{\ln{(2)}}{|g_{ee}|^2}\frac{\Lambda^2}{Q^{11}}\left[  \left(\frac{G_F}{\sqrt{2}}\right)^4 \left(\frac{1}{q^2}\right)^2 \left(\frac{ y_t^2 y_b^2 y_e^2 v^2}{(16\pi^2)^4}\right)^2 +  \left(\frac{G_F}{\sqrt{2}}\right)^2 \frac{1}{q^2} \left(\frac{y_t y_b y_e v}{(16\pi^2)^2\Lambda^2}\right)^2  + \frac{1}{\Lambda^8}\right]^{-1}.
\label{eq:halflife}
\end{equation}
The effective $Q$-value of the decay process can be extracted from analyses of the data from the KamLAND-Zen experiment  \cite{KamLAND-Zen:2016pfg} and turns out to be $\mathcal{O}(10\,{\rm MeV})$. The factor of $(1/q^2)$ comes from the neutrino propagator and is typically of order $100\,{\rm MeV}$, the inverse distance-scale between nucleons. Combining these, our estimate for the half-life as a function of $\Lambda$  is depicted in the left panel of Fig.\,\ref{fig:Lifetime}. For $\mathcal{O}(1)$ couplings, the current lifetime lower-bound -- $\Lambda\gtrsim5$~TeV -- and the neutrino mass requirements -- Eq.~(\ref{eq:nu_mass}) -- are incompatible. This strongly suggests that if there is new physics that manifests itself via $\mathcal{O}^{ee}_s$ at the tree level, this new physics is not responsible for generating the observed nonzero neutrino masses.

\begin{figure}[!t]
\centering 
\includegraphics[width=1\textwidth, height=0.5\textwidth]{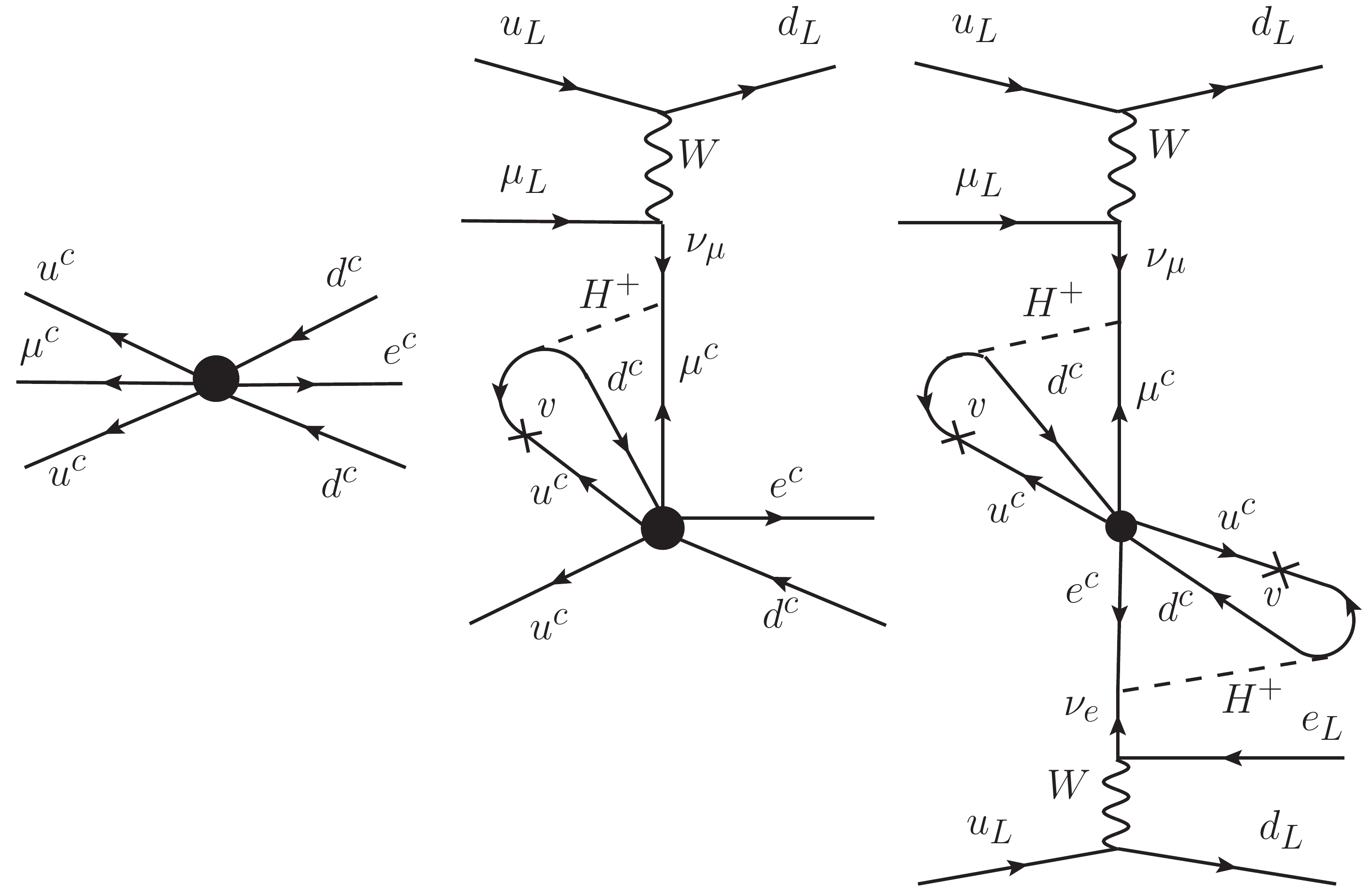}
\caption{Feynman diagrams contributing to $\mu^-\to e^+$-conversion from the dimension-nine all-singlets operator $\mathcal{O}^{\mu e}_s$ at the tree level (left), two-loop level (middle), and four-loop level (right).  The blob represents the effective operator while the $\times$ represents the Higgs-boson vacuum expectation value.}
\label{fig:Mutoe}
\end{figure}
Fig.~\ref{fig:Mutoe} depicts the tree-level, two-loop and four-loop contributions to $\mu^- \to e^+$-conversion from $\mathcal{O}^{e\mu}_s$. In order to estimate $R_{\mu^- e^+}$,  as defined in Eq.\,(\ref{mutoe}), we estimate the muon capture rate, as outlined in \cite{Berryman:2016slh}, to be
\begin{equation}
 \Gamma_{\mu^-}\propto \left(\frac{G_F}{\sqrt{2}}\right)^2 \left(\frac{Z_{\rm eff}^3}{\pi \left(a_0 m_e/m_\mu\right)^3}\right)Q^2\,,
 \label{eq:muoncapture}
\end{equation}
where $Z_{\rm eff}$ is the effective atomic number, $a_0$ the Bohr radius, and $Q$ the estimated typical energy of the process, of order the muon mass  $m_\mu$. While estimating $R_{\mu^- e^+}$, the term in the second parentheses in Eq.~(\ref{eq:muoncapture}) cancels out in the ratio, yielding
\begin{equation}
 R_{\mu^- e^+} = | g_{ e \mu } |^2 \frac{ Q^6 }{ \Lambda^2 } \left[ \left( \frac{ G_F }{ \sqrt{2} } \right)^2 \left(\frac{ 1 }{ q^2 }\right)^2 \left( \frac{ y_t^2 y_b^2 y_\mu y_e v^2 }{ ( 16 \pi^2 )^4 } \right)^2 + \frac{ 1 }{ q^2 } \left( \frac{ y_t y_b y_\mu v }{ ( 16 \pi^2 )^2 \Lambda^2 } \right)^2 + \left( \frac{ \sqrt{2} }{ G_F } \right)^2 \frac{ 1 }{ \Lambda^8 } \right].
 \label{eq:mutoe}
\end{equation}
 The normalized conversion rate for this process as a function of $\Lam$ is depicted in Fig.\,\ref{fig:Lifetime} along with the current bounds on the process from the SINDRUM~II collaboration \cite{Bertl:2006up}, and the expected {\it Mu2e} sensitivity, Eq.~(\ref{mu2ebound}). The current bound from SINDRUM~II implies that $\Lam \gtrsim 10\,{\rm GeV}$ for $\mathcal{O}(1)$ couplings. Again, the neutrino mass requirements are inconsistent with the existing $\mu^- \to e^+$-conversion bounds.

If all $g_{\alpha\beta}$ are of the same magnitude, current constraints on $\Lambda$ from $0\nu\beta\beta$ -- $\Lambda\gtrsim 1$~TeV for $g_{ee}$ of order one --  would translate into unobservable rates for $\mu^-\to e^+$-conversion in nuclei. However, there are no model-independent reasons to directly relate, e.g., $g_{\mu e}$ to $g_{ee}$, hence the bounds from $0\nu\beta\beta$ need not apply directly to searches for $\mu^-\to e^+$-conversion. Model-dependent considerations are required in order to explore possible relations between $g_{ee}$ and $g_{\mu e}$. On the the other hand, observable  rates for $\mu^-\to e^+$-conversion require $\Lambda\lesssim 100$~GeV and hence new particles with masses around (or below) the weak scale. It is natural to suspect that models associated to such small effective scales are also vulnerable to lepton-number conserving, low-energy observables, especially searches for CLFV.  As argued in the introduction, these phenomena can only be addressed within UV-complete models, which we introduce and discuss in the next section.

\setcounter{equation}{0}
\section{Ultraviolet completions of the effective operator $O^{\al\be}_s$}
\label{sec:UV}

\begin{figure}[t]
\centering 
\includegraphics[clip,width=0.8\linewidth]{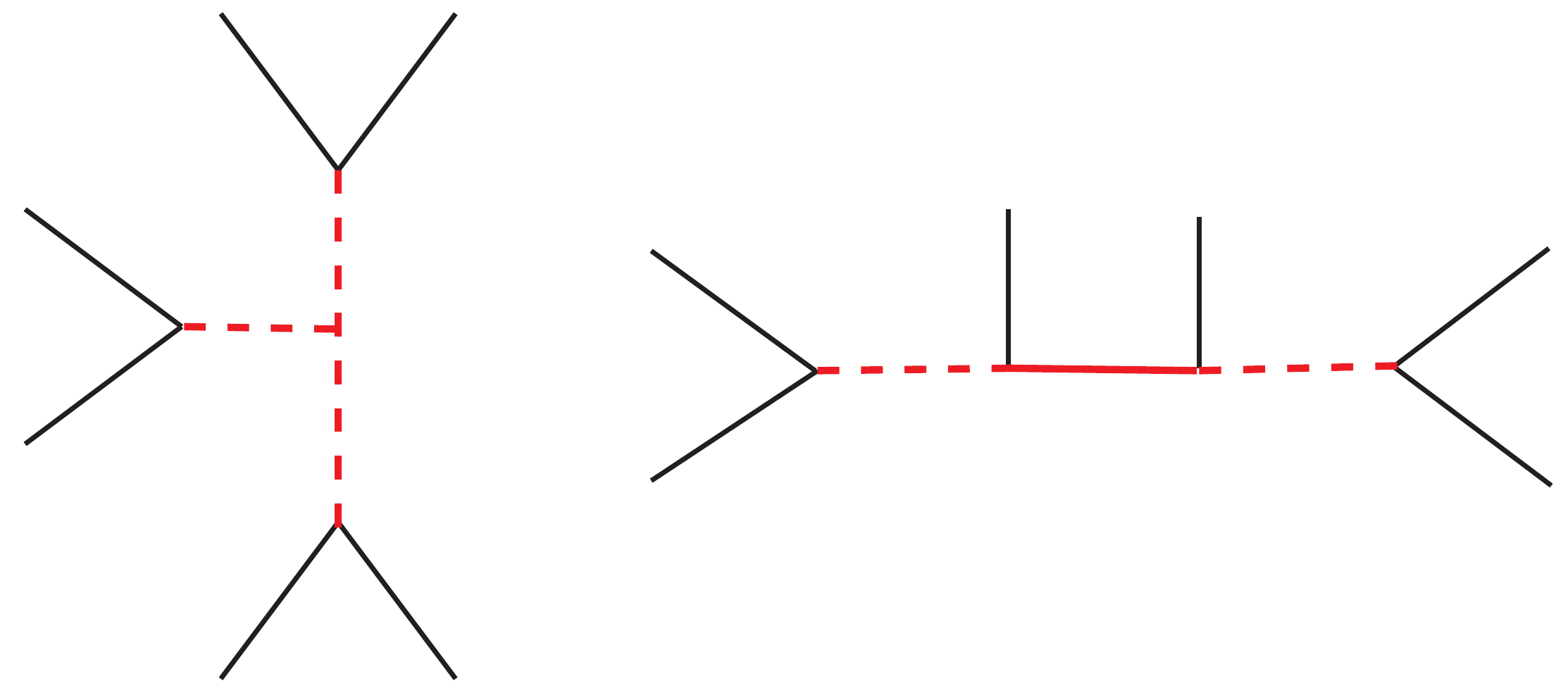}
\caption{Topologies that realize the all-singlets dimension-nine operator $\mathcal{O}_s$ at tree level. Topology 1 (left) involves only new bosons while Topology 2 (right) requires both new fermions and bosons. In both topologies, the bosons can be scalars or vectors.}
\label{fig:Topology}
\end{figure}

Here we discuss tree-level UV-completions of the all-singlets dimension-nine operator $\mathcal{O}^{\al\be}_s$, introduced in the previous section, Eq.~(\ref{eq:effectiveOperatorGen}). As all fields in the effective operator are fermions, all new interactions involving SM fields are either Yukawa or gauge interactions, i.e, they are all 3-point vertices. Furthermore, relevant interactions involving only new-physics fields are at most also 3-point vertices. This is due to the fact that the operator in question has six fermions in the final state and we are only interested in tree-level realizations of   $\mathcal{O}^{\al\be}_s$. Since only 3-point vertices are possible, there are only two topologies that lead to $\mathcal{O}^{\al\be}_s$ at the tree level \cite{Helo:2015fba,Anamiati:2018cuq}:
\begin{enumerate}
\item All new particles are bosons. Each boson couples to a pair of SM fermions, and three new-physics bosons define a new interaction vertex. This is depicted in the left panel of Fig. \ref{fig:Topology}. The new-physics bosons can be scalars or vectors.

\item All new interactions involve one boson and two fermions. The new particles are bosons and fermions and SM fields either couple pair-wise with a new-physics boson or couple to a new-physics boson and a new-physics fermion. This is depicted in the right panel of Fig.~\ref{fig:Topology}. Again, the new-physics bosons can be scalars or vectors.
\end{enumerate}

\begin{table}[!t]
\caption{Quantum numbers of all possible pairs of the $SU(2)_L$ gauge singlet Standard Model fermions $\ell^c,u^c,\overline{d^c}$.}
	\centering
	\begin{tabular}{|p{2.5cm}|p{8cm}|}
	\hline	&\\
	Pairs  (Lorentz) & Representation under $\bigl({\rm SU}(3)_{\rm C},\,{\rm SU}(2)_{\rm L}\bigr)_{{\rm U}(1)_{\rm Y}}$\\
	\hline	&\\
		$\ell^c \ell^c$ (scalar) & $(1,1)_1 \times (1,1)_1 = (1,1)_2$ \\
		$\ell^c u^c$ (scalar) & $(1,1)_1 \times (\overline{3},1)_{-2/3} = (\overline{3},1)_{1/3}$ \\
		$\ell^c \overline{d^c}$ (vector) & $(1,1)_1 \times (3,1)_{-1/3} = (3,1)_{2/3}$ \\
		$u^c u^c$ (scalar) & $(\overline{3},1)_{-2/3} \times (\overline{3},1)_{-2/3} =(3_a,1)_{-4/3}+ (\overline{6}_s,1)_{-4/3}$ \\
		$u^c \overline{d^c}$ (vector) & $(\overline{3},1)_{-2/3} \times (3,1)_{-1/3} = (1,1)_{-1} + (8,1)_{-1}$ \\
		$\overline{d^c}\, \overline{d^c}$ (scalar) & $(3,1)_{-1/3} \times (3,1)_{-1/3} = (\overline{3}_a,1)_{-2/3}+ (6_s,1)_{-2/3}$\\
		&\\
		\hline
	\end{tabular}
	\label{tab:pairs}
\end{table}
\begin{table}[!t]
\caption{Quantum numbers of all the possible triplets of $SU(2)_L$ gauge singlet Standard Model fermions $\ell^c,u^c,\overline{d^c}$ (with at most two identical fields).}
	\centering
	\begin{tabular}{|p{1.3cm}|p{13.cm}|}
                 \hline    & \\
                 Triplets  & Representation under $\bigl({\rm SU}(3)_{\rm C},\,{\rm SU}(2)_{\rm L}\bigr)_{{\rm U}(1)_{\rm Y}}$\\
	\hline	&\\
		$\ell^c \ell^c u^c$ & $(1,1)_1 \times (1,1)_1 \times (\overline{3},1)_{-2/3} = (\overline{3},1)_{4/3}$ \\
		$\ell^c \ell^c \overline{d^c}$ & $(1,1)_1 \times (1,1)_1 \times (3,1)_{-1/3} = (3,1)_{5/3}$ \\
		$\ell^c u^c u^c$ & $(1,1)_1 \times (\overline{3},1)_{-2/3} \times (\overline{3},1)_{-2/3} = (3_a, 1)_{-1/3} + (\overline{6}_s,1)_{-1/3}$ \\
		$\ell^c \overline{d^c}\, \overline{d^c}$ & $(1,1)_1 \times (3,1)_{-1/3} \times (3,1)_{-1/3} = (\overline{3}_a,1)_{1/3} + (6_s,1)_{1/3}$ \\
		$\ell^c u^c \overline{d^c}$ & $(1,1)_1 \times (\overline{3},1)_{-2/3} \times (3,1)_{-1/3} = (1,1)_0 + (8,1)_0$ \\
		$u^c u^c \overline{d^c}$ & $(\overline{3},1)_{-2/3} \times (\overline{3},1)_{-2/3} \times (3,1)_{-1/3} = \left[ (3_a,1)_{-4/3} + (\overline{6}_s,1)_{-4/3} \right] \times (3,1)_{-1/3} = (\overline{3},1)_{-5/3} + (6,1)_{-5/3} + (\overline{3},1)_{-5/3} + (\overline{15},1)_{-5/3}$ \\
		$u^c \overline{d^c}\, \overline{d^c}$ & $(\overline{3},1)_{-2/3} \times (3,1)_{-1/3} \times (3,1)_{-1/3} = (\overline{3},1)_{-2/3} \times \left[ (\overline{3}_a,1)_{-2/3} + (6_s,1)_{-2/3} \right] = (3,1)_{-4/3} + (\overline{6},1)_{-4/3} + (3,1)_{-4/3} + (15,1)_{-4/3}$\\
		&\\
		\hline
	\end{tabular}
	\label{tab:triples}
\end{table}

In order to systematically analyze the different internal particles that can appear in Fig.~\ref{fig:Topology}, we determine the quantum numbers of pairs and triplets of the external SM fermions of interest. The different combinations of pairs of fermions determine the possible quantum numbers of the bosons in the internal lines in Fig.~\ref{fig:Topology}. Similarly, different combinations of triplets of fermions determine the quantum numbers of the potential new fermions in the internal fermion line in Fig.~\ref{fig:Topology}(right).  

Table \ref{tab:pairs} lists all possible ways of pairing up any two $SU(2)_L$-singlet SM fermions.\footnote{Excluding left-handed antineutrinos $\nu^c$. We will comment on those later in this section.} Generation indices, for both leptons and quarks, have been omitted. Topology 1 can be realized by choosing three bosons with the same quantum numbers as these pairs, keeping in mind that there are two fermions of each type -- $u^c,d^c,\ell^c$ -- in $\mathcal{O}^{\al\be}_s$. For bilinear combinations of the same generation of quarks, only products symmetric in the color indices, i.e., forming a $6$ or $\overline{6}$ of $SU(3)_c$, exist since, e.g., $(u^c)^{\alpha i} (u^c)_{\alpha}^j = - (u^c)_{\alpha}^i (u^c)^{\alpha j} = (u^c)^{\alpha j} (u^c)_{\alpha}^i$, where $\alpha$ is the dummy Lorentz index and $(i,j)$ are the $SU(3)_c$ indices. Here, we will be concentrating on new-physics involving first-generation quarks, as we are interested in models that mediate $\mu^-\to e^+$-conversion at the tree level (left panel of Fig.~\ref{fig:Mutoe}). 
Unless otherwise noted, we will not consider models that ``mix'' different generations of the same quark-flavor. 

There are five different ``minimal'' realizations of Topology 1. Two of them involve heavy scalar bosons only, while the remaining three require new-physics vector and scalar bosons. Of course, one can consider ``less-minimal'' scenarios where one includes bosons with different quantum numbers associated to the same fermion-pair, e.g., the combination $u^c \overline{d^c}$ can connect to vector bosons in two different $SU(3)_c$ representations.

Similarly, Table \ref{tab:triples} lists all possible combinations of three $SU(2)_L$-singlet SM fermions. The different new-physics fermions that can make up Topology 2 must have the same quantum numbers as the combinations listed in the table. This list is exhaustive, and to get all possible diagrams, one needs to consider all allowed, distinct permutations of the triplets. In order to realize Topology 2, for each such combination, one needs to consider the possible ways of arranging fermion pairs, listed in Table \ref{tab:pairs}. It can be shown that this yields eighteen different ``minimal'' realizations of Topology 2, not considering the different representations for the same combination of SM fermions. 

Next, we want to ensure that, at the tree level, the different new-physics scenarios lead to the all-singlets operator but not to other dimension-nine (or lower dimensional) LNV operators. New particles with the same quantum numbers as some of the combinations in Table \ref{tab:pairs} can also couple to pairs of SM fermions that contain the $SU(2)_L$-doublets $L,Q$. For example, the pair $\ell^c u^c$ transforms like a $(\overline{3},1)_{1/3}$. A scalar that couples to this pair of SM fermions can also couple to $\overline{L}\, \overline{Q}$, since the
latter has identical quantum numbers. These new bosons would lead to, along with  the all-singlets operator, other six-fermion operators, including $(\overline{L} \,\overline{Q})(\overline{L}\, \overline{Q})(\overline{d^c}\, \overline{d^c})$ (for a complete list, see Tables I, II and III  in \cite{Berryman:2016slh}). Unlike the all-singlets operator, all other dimension-nine operators saturate the constraints associated to non-zero neutrino masses for $\Lambda$ values that translate into tiny rates for $\mu^-\to e^+$-conversion, see Figure 7 in \cite{Berryman:2016slh}. 

In order to systematically address this issue, we list all the relevant SM fermion pairs that transform in the same way in Table~\ref{tab:smfermionstransformingidentically}. The pairs relevant for $\mathcal{O}^{\al\be}_s$ are shown in red. From the table, one can see that a new particle that couples to, e.g., $\ell^c$ with $u^c$ or $\overline{d^c}$ can also couple to $\overline{L}\, \overline{Q}$, and so on. The table reveals that there are two avenues for avoiding unwanted couplings. One is to have one of the new bosons couple to the pair $\ell^c\ell^c$, which is not degenerate, quantum-number-wise, with any other pair of SM fermions. The other is to add a new fermion and a new boson such that $\ell^c$ couples to them in Topology 2. The reason for this is that all other pairings involving $\ell^c$ have an unwanted ``match,'' see Table~\ref{tab:smfermionstransformingidentically}. This extra requirement drastically reduces the total number of minimal models for the two topologies, and allows us to write down all possible UV completions with no more than three new particles. The final allowed combinations and the corresponding new particles are listed in Table~\ref{tab:finallist}. The list is exhaustive, and all possible UV completions of $\mathcal{O}^{\al\be}_s$ at the tree level can be implemented with a subset of less than or equal to three of these particles.

It is also important to consider whether new interactions would materialize if neutrino $SU(2)_L$-singlet fields, $\nu^c$, were also present. Pairings that include $\nu^c$ are also included in  
Table~\ref{tab:smfermionstransformingidentically}. Given all constraints discussed above, there are no new couplings involving $\nu^c$ other than the neutrino Yukawa coupling and $\nu^c$ Majorana masses for new-physics models that do not contain the vector $C^\mu \sim (1,1)_1$ field. In models that contain $C^\mu$, one need also consider the interaction term  $\overline{\ell^c}\overline{\sigma}^\mu \nu^c_i C_\mu$. We return to the left-handed antineutrinos and the mechanism behind neutrino masses in Sec.~\ref{sec:conclusion}.
\begin{table}[!t]
\caption{Pairs of Standard Model fermions that share the same gauge quantum numbers. The pairs of interest here are in red. The pair $\ell^c\ell^c$ does not transform like any other pair of SM fields; the same is true of the color-symmetric pairs of $u^cu^c$ and $d^cd^c$. }
	\centering
	\begin{tabular}{|p{5cm}|p{5cm}|}
	\hline  &  \\
	Fermion pairs transforming as  & $\bigl({\rm SU}(3)_{\rm C},\,{\rm SU}(2)_{\rm L}\bigr)_{{\rm U}(1)_{\rm Y}}$\\
	\hline	&\\
    $LL$, $\overline{\ell^c}\, \overline{\nu^c}$ & $(1,1)_{-1}$ scalar\\
    $\color{red} \overline{d^c} u^c$, $\overline{\ell^c} \nu^c$ & $(1,1)_{-1}$ vector\\
		$\color{red} \ell^c u^c$, $\overline{u^c} \overline{d^c}$, $Q^2$, $\overline{L}\, \overline{Q}$, $d^c \nu^c$ & $(\overline{3},1)_{1/3}$ scalar\\
		$\overline{u^c} \overline{d^c}$, $QQ$ & $(6,1)_{1/3}$ scalar \\
    $\color{red} d^c d^c$, $\overline{u^c}\, \overline{\nu^c}$ & $(3,1)_{2/3}$ scalar\\
		$\color{red} \overline{d^c} \ell^c$, $\overline{L} Q$, $\overline{u^c} \nu^c$ & $(3,1)_{2/3}$ vector \\
		$\color{red} u^c u^c$, $\overline{d^c}\, \overline{\ell^c}$ & $(3,1)_{-4/3}$ scalar \\
    $\nu^c \nu^c$, $\overline{\nu^c} \,\overline{\nu^c}$ & $(1,1)_0$ scalar\\
		$L \overline{L}$, $Q \overline{Q}$, $\ell^c \overline{\ell^c}$, $d^c \overline{d^c}$, $u^c \overline{u^c}$, $\overline{\nu^c} \nu^c$ & $(1,1)_0$ vector \\
		$Q \overline{Q}$, $d^c \overline{d^c}$, $u^c \overline{u^c}$ & $(8,1)_0$ vector\\
		&\\
		\hline
	\end{tabular}
	\label{tab:smfermionstransformingidentically}
\end{table}
\begin{table}[!t]
\caption{All new particles required for all different tree-level realizations of the all-singlets dimension-nine operator $\mathcal{O}^{\al\be}_s$, according to the restrictions discussed in the text. All particles are $SU(2)_L$ singlets. The fermions $\psi$, $\zeta$, and $\chi$ come with a partner ($\psi^c$, $\zeta^c$, and $\chi^c$ respectively), not listed. We don't consider fields that would couple to the antisymmetric combination of same-flavor quarks since these cannot couple quarks of the same generation.}
	\centering
	\begin{tabular}{|p{4cm}|p{4cm}|p{4cm}|}
	\hline  
	& & \\
	New particles   & $\,\,\bigl({\rm SU}(3)_{\rm C},\,{\rm SU}(2)_{\rm L}\bigr)_{{\rm U}(1)_{\rm Y}}$ & $\,\,$Spin\\
	\hline	& &\\
		$\Phi\equiv (\overline{l^c}\,\overline{l^c})~$ & $(1,1)_{-2}$ & $\,\,$ scalar \\
		$\Sigma\equiv (\overline{u^c}\,\overline{u^c})$ & $(6,1)_{4/3}$ & $\,\,$ scalar\\
		$\Delta \equiv (\overline{d^c}\,\overline{d^c})$ & $(6,1)_{-2/3}$&  $\,\,$ scalar\\
		$ C \equiv (\overline{u^c}\,d^c)$ & $(1,1)_{1},\,\,\,(8,1)_{1}$& $\,\,$ vector\\
		$ \psi \equiv (u^c\,l^c\,l^c) $ & $(\overline{3},1)_{4/3}$& $\,\,$ fermion\\
		$ \zeta \equiv (d^c\,\overline{l^c}\,\overline{l^c})$ & $(\overline{3},1)_{-5/3}$&$\,\,$ fermion\\
		$ \chi \equiv (l^c\,u^c\,u^c)$ & $(\overline{6},1)_{-1/3}$&$\,\,$ fermion\\
		$ N \equiv (l^c\,\overline{d^c}\,u^c)$ & $(1,1)_{0}$,\,\,\,$(8,1)_{0}$&$\,\,$ fermion \\
		& &\\
		\hline
	\end{tabular}
	\label{tab:finallist}
\end{table}

In the following subsections we list all the different models. We divide them into different categories. Some models contain new vector bosons, others contain only new-physics scalars or fermions. Since all new particles need to be heavy, including potential new vector bosons, no-vectors models are easier to analyze since, as is well-known, consistent quantum field theories with massive vector bosons require extra care. There are, altogether, eight models: four with and four without new massive vector fields. We discuss the no-vectors models first. We will also broadly distinguish models based on whether they also lead to the violation of baryon-number conservation and whether any flavor-structure naturally arises.

\subsection{No-vectors Models}
Here, all no-vectors models are discussed in turn. Models are named according to the new-physics field content, see Table~\ref{tab:finallist}. Explicitly, they are (1) $\zeta\Phi\Sigma$, (2) $\chi\Delta\Sigma$, (3) $\psi\Delta\Phi$, and (4) $\Phi \Sigma \Delta$. The first three realize ${\cal O}^{\alpha\beta}_s$ via topology 2 (Fig.~\ref{fig:Topology}(right)) while the last one realizes ${\cal O}_s^{\alpha\beta}$ via topology 1 (Fig.~\ref{fig:Topology}(left)). 

\subsubsection{Model $\zeta\Phi\Sigma$}
Here, the SM particle content is augmented by a couple of vector-like fermions $\zeta\equiv (\overline{3},1)_{-5/3}$ and $\zeta^c\equiv (3,1)_{5/3}$, the color-singlet scalar $\Phi \sim (1,1)_{-2}$, and the colored scalar $\Sigma \sim (6,1)_{4/3}$. The most general renormalizable Lagrangian is 
\begin{equation}
	\mathcal{L}_{\zeta\Phi\Sigma} = \mathcal{L}_{\rm SM} + \mathcal{L}_{\rm kin}  + y_{\Phi\al \be}\, \Phi \ell^c_\al \ell^c_\be +  y_{\Sigma u}\,\Sigma u^c u^c + y_{\Phi\zeta^c}\, \Phi \zeta^c d^c +  y_{\Sigma\zeta}\, \Sigma \zeta d^c + m_{\zeta}\, \zeta \zeta^c + V(\Phi,\Sigma,0) + {\rm h.c.}\,,
	\label{eq:lagrangianModel1}
\end{equation}
where $ \mathcal{L}_{\rm SM}$ is the SM Lagrangian, $\mathcal{L}_{\rm kin}$ contains the kinetic-energy terms for the new particles, and $V(\Phi,\Sigma,0)$ is the most general scalar potential involving the scalars $\Phi,\Sigma$, written out explicitly in Appendix~\ref{app:potential}. By design, lepton number is violated by two units but it is conserved in the limit where any of the new Yukawa couplings vanishes. On the other hand, baryon number is conserved. In units where the quarks have baryon-number one, $\Sigma$ can be assigned baryon-number $+2$, $\zeta,\zeta^c$ baryon-number $-1,+1$, respectively, and $\Phi$ baryon-number zero. 

It is easy to check that this model realizes ${\cal O}^{\alpha\beta}_s$ via topology 2 (Fig.~\ref{fig:Topology}(right)) and
\begin{equation}
 \frac{g_{\alpha\beta}}{\Lambda^5}\equiv \frac{y_{\Phi \alpha\beta}\,y_{\Phi\zeta^c}^*\,y_{\Sigma\zeta}^*\,y_{\Sigma u}}{M_\Phi^2\,M_\Sigma^2\,m_\zeta}\,.
 \label{Mod1:MutoE}
\end{equation}
Here, $y_{\Phi \alpha\beta}$ controls the lepton-flavor structure of the model. $\mu^-\to e^+$-conversion rates are proportional to $|y_{\Phi e\mu}|^2$, while those for $0\nu\beta\beta$ are proportional to $|y_{\Phi ee}|^2$.

The new-physics states will also mediate CLFV phenomena, sometimes at the tree level. In what follows, we write down the effective operators that give rise to different CLFV processes, and estimate bounds on the effective scales of these operators. The CLFV observables of interest are:
\begin{enumerate}
\item  $\mu^{\pm}\rightarrow e^{\pm} e^{\pm} e^{\mp}$ decay: The effective Lagrangian giving rise to this decay, generated at the tree level, is
 \begin{equation}
  \mathcal{L}_{\mu\to 3e} = \frac{y_{\Phi e\mu}\,y_{\Phi ee}^*}{M_\Phi^2}\,\left( \mu^c e^c\right)\,\left( \overline{e^c}\, \overline{e^c}\right)\,,
 \end{equation}
 and the relevant Feynman diagram is depicted in the left panel of Fig.\,\ref{fig:NV1_CLFV}. 
 The strongest bounds on  $\mu^{+}\rightarrow e^{+} e^{-} e^{+}$ come from the SINDRUM spectrometer experiment  \cite{Bellgardt:1987du}:
 \begin{equation}
  {\rm Br}(\mu^{+}\rightarrow e^{+} e^{-} e^{+})< 1.0\times 10^{-12}\,.
  \label{eq:mu3e}
 \end{equation}
 Assuming the phase-space distributions are similar to those of ordinary $\mu$-decay $(\mu\rightarrow e\overline{\nu}_e\nu_\mu)$, this translates into \cite{Kuno:1999jp,Fael:2016yle}
 \begin{equation}
  \frac{|y_{\Phi e\mu}\,y_{\Phi ee}^*|^2}{M_\Phi^4}\leq 1.4\times 10^{-22} \,{\rm GeV}^{-4}\,,
 \end{equation}
or $M_\Phi\geq 290\,{\rm TeV}$ for $\mathcal{O}(1)$ couplings. The Mu3e experiment, under construction at PSI,  aims to reach sensitivities better than $10^{-15}$ on this channel \cite{Berger:2014vba} and hence sensitivity to $\Phi$-masses around $1000~{\rm TeV}$ \cite{deGouvea:2013zba}. 

\item $\mu^+ \to e^+ \gamma$ decay: At the one-loop level, $\Phi$-exchange also mediates, as depicted in the middle panel of Fig.\,\ref{fig:NV1_CLFV}, $\mu^+\to e^+\gamma$. The effective operator governing $\mu \to e \, \ga$ is 
\begin{equation}
   \mathcal{L}_{\mu^+\to e^+\ga} = \frac{ y_{\Phi\mu\mu}^*\,y_{\Phi\mu e}\,(2e)\,y_\mu }{16 \pi^2M_\Phi^2} (L \overline{H})\,\sigma^{\alpha\beta}e^c F_{\alpha\beta}\,, 
 \end{equation} 
where $y_{\mu}$ is the muon Yukawa coupling. 
Experimentally, the most stringent constraints come from the  MEG experiment at PSI~\cite{TheMEG:2016wtm} 
\begin{equation}
 {\rm Br}(\mu^+ \to e^+ \gamma)=4.2\times 10^{-13}\,.
\end{equation} 
Using results from \cite{Raidal:1997hq}, we get
\begin{align}
\text{Br}(\mu^+ \to e^+  \ga) \approx 5.3 \times 10^{-6}  \frac{|y_{\Phi\mu\mu}^*\,y_{\Phi\mu e}|^2}{M_\Phi^4~(\text{TeV})} < 4.2 \times 10^{-13} \, ,
\label{eq:mutoegamma}
\end{align}
which leads to $M_\Phi \gtrsim 60$~TeV, given $\mathcal{O}(1)$ couplings. As expected, the $\mu^+ \to e^+ \ga$ bound is weaker than that of $\mu \to 3\, e$ as the former is loop-suppressed.
The upgraded MEG-II experiment plans to reach a sensitivity of $10^{-14}$ with three years of data taking \cite{Baldini:2013ke}.
\begin{figure}[!t]
\centering 
\includegraphics[width=0.55\textwidth, height=0.25\textwidth]{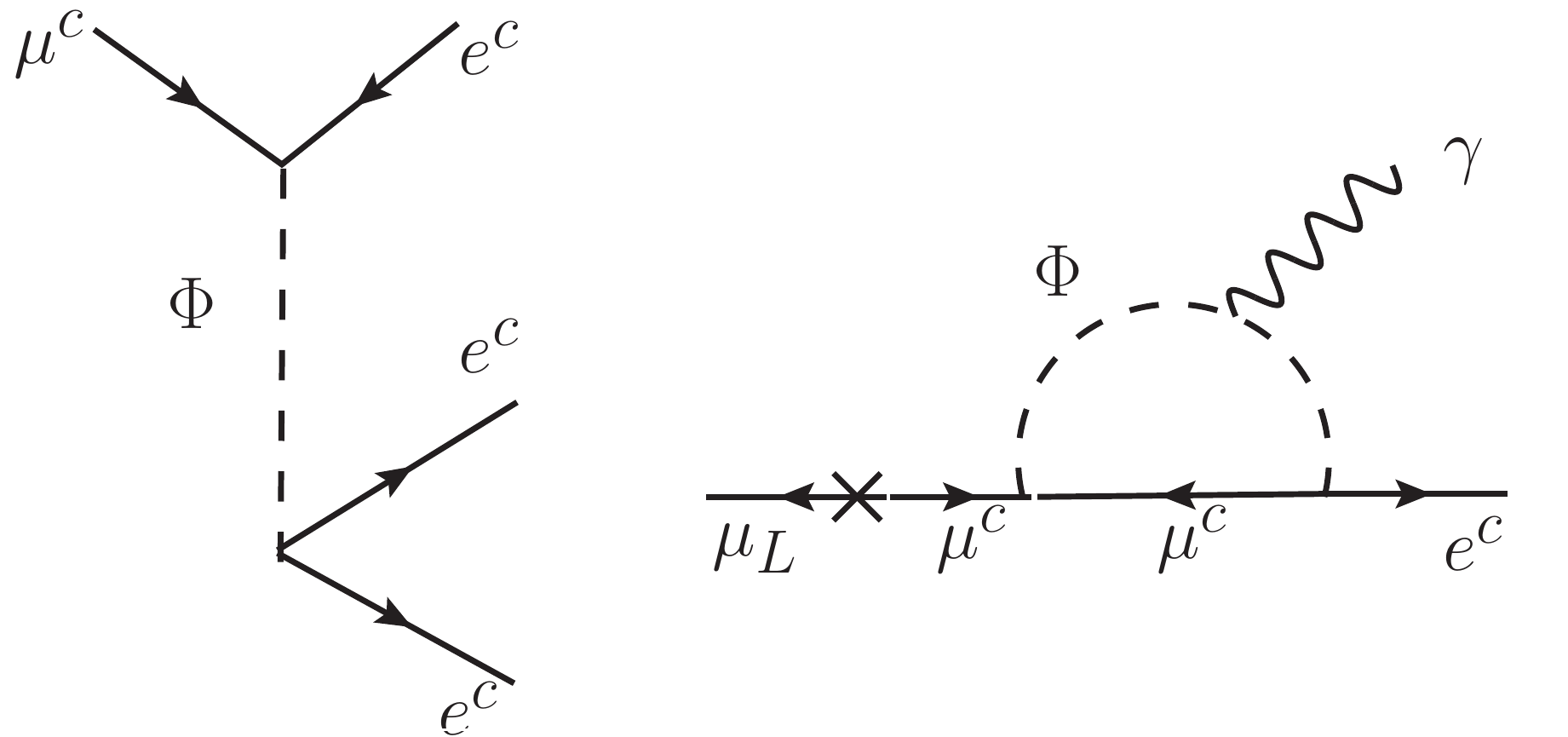}\,\,\includegraphics[width=0.35\textwidth, height=0.2\textwidth]{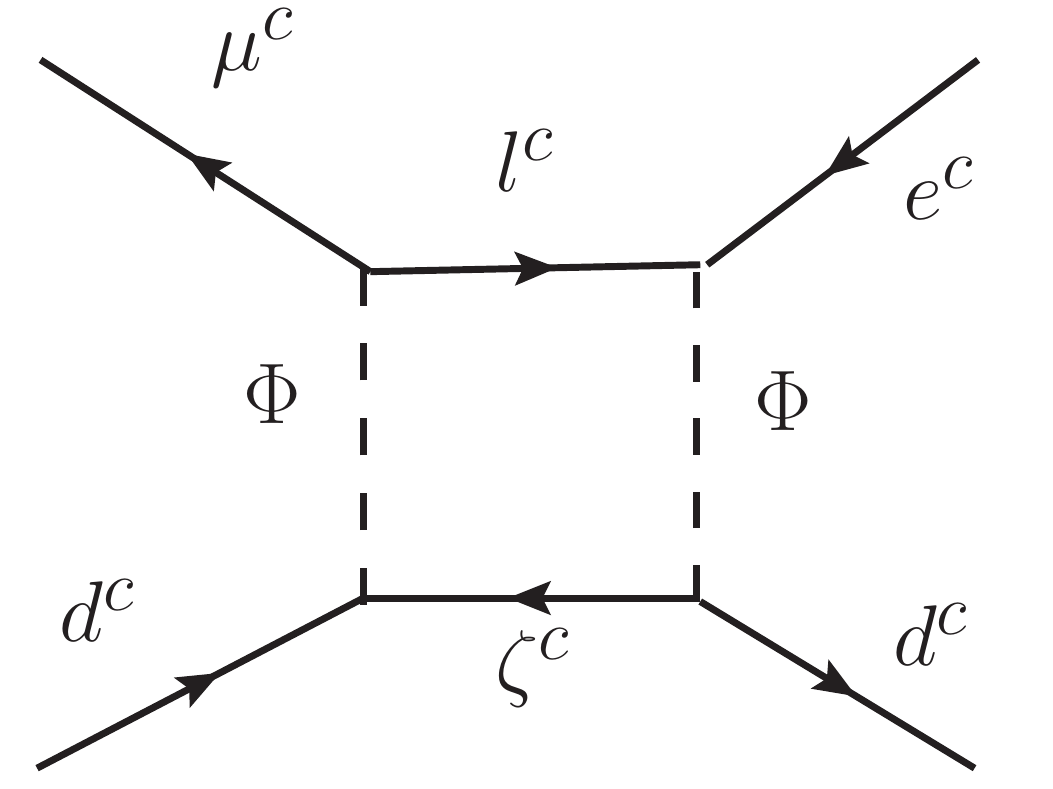}
\caption{Feynman diagrams contributing to the CLFV processes $\mu\to 3e$ (left), $\mu^+ \to e^+ \gamma$ (middle), and $\mu^- \to e^-$-conversion (right) in Model $\zeta\Phi\Sigma$.}
\label{fig:NV1_CLFV}
\end{figure}
\item $\mu^-\to e^-$-conversion in nuclei: In this model, $\mu^-\to e^-$-conversion occurs at the one-loop level, as depicted in the right panel of Fig.\,\ref{fig:NV1_CLFV}. The effective Lagrangian can be estimated as
\begin{equation}
  \mathcal{L}_{\mu\to e} = \frac{y_{\Phi \mu\beta}^*\,y_{\Phi e\beta}\,y_{\Phi\zeta^c}\,y_{\Phi\zeta^c}^*}{16\pi^2 \Lambda^2}\, \left(\overline{\mu^c} e^c\right) \left(\overline{d^c} d^c\right)\,,
\label{Eq:MuMtoEM}
\end{equation}
where $\Lambda$ is the effective scale, a function of $M_\Phi$ and $M_\zeta$, and the lepton-index $\beta$ is summed over. Note that this operator is also sensitive to the $y_{\Phi e\tau}$ and $y_{\Phi \mu\tau}$ couplings. An extra contribution comes from the middle panel of Fig.\,\ref{fig:NV1_CLFV}, where the photon is put offshell, and radiates a $q\overline{q}$ pair.

The SINDRUM II experiment at PSI constrains $\mu^-\to e^-$-conversion in gold \cite{Bertl:2006up}:
\begin{equation}
R_{\mu^- e^-}^\text{Au} \equiv \frac{\Gamma(\mu^- + \text{Au} \to e^- + \text{Au})}{\Gamma(\mu^- + \text{Au} \to \nu_\mu + \text{Pt})} < 7 \times 10^{-13} \text{ (90\% CL).}
\end{equation}
Using Eq.\,(\ref{eq:muoncapture}), we estimate \cite{Kuno:1999jp}
\begin{equation}
 R_{\mu^- e^-}= \left|\frac{\sqrt{2}}{G_F}\, \frac{y_{\Phi \mu\beta}^*\,y_{\Phi e\beta}\,y_{\Phi\zeta^c}\,y_{\Phi\zeta^c}^*}{16\pi^2 \Lambda^2}\right|^2.
\end{equation}
For $\mathcal{O}(1)$ couplings, this yields $\Lambda\geq 30\,{\rm TeV}$. Stronger sensitivity is expected from the next-generation experiments COMET \cite{Kuno:2013mha}, DeeMe \cite{Natori:2014yba}, and {\it Mu2e} \cite{Bartoszek:2014mya}, as discussed in the introduction. Ultimately, one would be sensitive to $\Lambda$ scales up to a few 100~TeV.

\item Muonium-Antimuonium oscillations $(\mu^+ e^- \rightarrow  \mu^- e^+)$:
Muonium $({\rm Mu})$ is the bound state of an $e^-$ and a $\mu^+$, whereas its anti-partner, the antimuonium $(\overline{\rm Mu})$ is the bound state of an $e^+$ and a $\mu^-$. Muonium-antimuonium oscillation is a process where muonium converts to antimuonium, thereby changing both electron-number and muon-number by two units \cite{Kuno:1999jp}. Here, the effective Lagrangian governing this process at the tree level is
 \begin{equation}
  \mathcal{L}_{{\rm Mu}-\overline{\rm Mu}} = \frac{y_{\Phi \mu\mu}\,y_{\Phi ee}^*}{M_\Phi^2}\,\left(\mu^c \mu^c \right)\,\left(\overline{e^c}\, \overline{e^c}\right)\,.
 \end{equation}
 The probability that a Mu bound state at $t=0$ is detected as a $\overline{\rm Mu}$ bound state at a later time is proportional to
 $\left(y_{\Phi \mu\mu}\,y_{\Phi ee}^*\right)/M_\Phi^2$. The upper limit quoted by the PSI
 experiment \cite{PhysRevLett.82.49} yields
 $ \left(y_{\Phi \mu\mu}\,y_{\Phi ee}^*\right)/M_\Phi^2\lesssim 0.002\, G_F$ or
 \begin{equation}
  \frac{|y_{\Phi \mu\mu}\,y_{\Phi ee}^*|}{M_\Phi^2}\leq 2.5\times10^{-8}\,{\rm GeV}^{-2}\,,
  \label{eq:mod1MuMubar}
\end{equation}
which implies $M_\Phi\geq 6.3\,{\rm TeV}$ for $\mathcal{O}(1)$ couplings.

 \item Lepton--lepton scattering: $\Phi$-exchange will also mediate intermediate and high-energy scattering processes including $e^\pm\mu^\pm \rightarrow e^\pm \mu^\pm$, $e^+e^-\to e^+e^-$, and $e^+e^-\to\mu^+\mu^-$. If $M_{\Phi}$ is much larger than the center-of-mass-energies of interest, the following tree-level effective Lagrangian applies:
 \begin{equation}
  \mathcal{L}_{e\mu} = \frac{y_{\Phi e\mu}\,y_{\Phi e\mu}^*}{M_\Phi^2}\,e^c \mu^c \overline{e^c} \,\overline{\mu^c}
  \label{eq:mod1MueSc}
 \end{equation}

Measurements of $\sigma(e^+ e^- \to \mu^+ \mu^-)$ and $\sigma(e^+ e^- \to e^+ e^-)$  at LEP~\cite{LEP:2003aa} can be translated into constraints on the effective scale of the operator above,  
\begin{align} 
\frac{y_{\Phi e\mu}\,y_{\Phi e\mu}^*}{2 \, M_\Phi^2} \leq \frac{4\pi}{\Lambda^2_{\mu}} \;\; \text{and} \;\;
\frac{y_{\Phi ee}\,y_{\Phi ee}^*}{2 \, M_\Phi^2} \leq \frac{4\pi}{2 \,\Lambda^2_{e}},
\label{eq:scatt}
\end{align}
 where $\Lambda_{\mu} \approx 9.3$ TeV and $\Lambda_{e} \approx 8.9$ TeV. These translate into $M_\Phi \gtrsim 2.5$ TeV, given $\mathcal{O}(1)$ couplings.

 \item Anomalous magnetic moments: There is a well-known  discrepancy between the experimental value~\cite{Bennett:2006fi} and the SM prediction~\cite{Davier:2010nc,Hagiwara:2011af} of the anomalous magnetic moment of the muon, $ 10.1 \times 10^{-10} <a^{\text{exp}}_{\mu} - a^{\text{SM}}_{\mu}<  42.1\times 10^{-10}$ at the $2\sigma$ level.
The doubly charged $\Phi$-scalar will contribute to the muon $(g-2)$ at the one-loop level. The corresponding Feynman diagrams are quite similar
 to the middle panel of Fig.~\ref{fig:NV1_CLFV} with the external electron replaced by a muon. In the limit $\Phi$ is much heavier than muons and electrons, the resulting contribution is~\cite{Chakrabarty:2018qtt}~(see also~\cite{Moore:1984eg,Lindner:2016bgg})
 \begin{align}
 \Delta a_\mu = - \frac{ m_\mu^2 \left( y_{\Phi e\mu}\,y_{\Phi e\mu}^* + y_{\Phi \mu\mu}\,y_{\Phi \mu\mu}^* \right) }{ 6 \, \pi^2 M_\Phi^2 } \, .
 \end{align}
The negative sign of the contribution indicates that this type of new physics will not help alleviate the discrepancy. We can, nonetheless, derive a limit from the $g-2$ measurement by
requiring the absolute value of the contribution to be less than the discrepancy, which leads to $M_\Phi \gtrsim 734$ GeV, given the $\mathcal{O}(1)$ couplings. This bound is weaker than most of the previous ones discussed here. The Muon $g-2$ experiment, currently taking data at Fermilab, is ultimately expected to improve on the uncertainty of the muon $g-2$ by roughly a factor of two \cite{Grange:2015fou}.
 \end{enumerate}
 
A subset of the bounds estimated here is summarized in Fig.~\ref{fig:summaryBounds}. Not surprisingly, if all couplings of interest are of order one, constraints from $\mu\to 3e$ are the strongest and translate into $M_{\Phi}$ values that exceed hundreds of TeV. CLFV observables do not constrain, directly, $m_{\zeta}$ or $M_{\Sigma}$, while searches for $\mu^-\to e^-$-conversion are sensitive to both $M_{\Phi}$ and $m_{\zeta}$. Since both $\zeta$ and $\Sigma$ are colored, we expect LHC searches for exotic fermions or scalars to constrain, conservatively, $m_{\zeta},M_{\Sigma}\gtrsim$~500~GeV. We return to this issue briefly in Sec.~\ref{sec:coll}. Putting it all together, if all new-physics couplings are of order one, searches for CLFV imply upper bounds on the rate for $\mu^-\to e^+$-conversion that are much stronger than the sensitivity of next-generation experiments.  
 
Most of the CLFV bounds can be avoided, along with those from $0\nu\beta\beta$, if the flavor-structure of the new physics is not generic. In particular, in the limit where $y_{\Phi e\mu}$ is much larger than all other $y_{\Phi \alpha\beta}$ couplings, most of the constraints above become much weaker. This can be understood by noting that $\mu^-\to e^+$-conversion preserves an  $L_\mu - L_e$ (muon-number minus electron-number) global symmetry while the physics processes $\mu \rightarrow 3e$,  $\mu \rightarrow e \gamma$, $\mu^\pm \rightarrow e^\pm$-conversion, and $0\nu\beta\beta$ all violate $L_\mu - L_e$ by two units, while ${\rm Mu} - \overline{\rm Mu}$-oscillations violate $L_\mu - L_e$ by four units. In other words, if only the $\Phi \mu^c e^c$-coupling $y_{\Phi \mu e}$ is nonzero, the new-physics portion of the Lagrangian respects an $L_\mu - L_e$ global symmetry and all CLFV bounds vanish to a very good approximation.  The flavor-diagonal constraints from LEP and the muon $g-2$ do, however, apply, but are of order 1~TeV for $y_{\Phi \mu e}$ of order one, much less severe. This is a property of all new-physics scenarios that contain the $\Phi$-field since, in these scenarios, the only coupling of the leptons to the new degrees-of-freedom is the one to $\Phi$.

\subsubsection{Model $\chi\Delta\Sigma$}
Here, the SM particle content is augmented by a couple of vector-like fermions  $\chi   \sim (\overline{6},1)_{-1/3}$ and $\chi^c \sim
(6,1)_{1/3}$, and two colored scalars $\Sigma \sim (6,1)_{4/3}$ and $\Delta \sim
(6,1)_{-2/3}$. The most general renormalizable Lagrangian is
\begin{align}
 \mathcal{L}_{\chi\Delta\Sigma} = &\, \mathcal{L}_{\rm SM} + \mathcal{L}_{\rm kin} +  y_{\Sigma u}\,\Sigma u^c u^c + y_{\Delta d}\,\Delta d^c d^c + y_{\Delta\chi}\, \Delta \overline{\chi} \,\overline{\chi} + y_{\Delta\chi^c}\,\Delta \chi^c \chi^c + y_{\Sigma \alpha}\, \overline{\Sigma} \chi^c \ell_{\alpha}^c + y_{\Delta \alpha}\, \Delta \chi \ell_{\alpha}^c \nonumber\\&+   m_{\chi}\,\chi\chi^c + V(0,\Sigma,\Delta) + {\rm h.c.}\,,
\label{eq:lagrangianModel5}
\end{align}
where $ \mathcal{L}_{\rm SM}$ is the SM Lagrangian, $\mathcal{L}_{\rm kin}$ contains the kinetic-energy terms for the new particles, and $V(0,\Sigma,\Delta)$ is the most general scalar potential involving the scalars $\Delta,\Sigma$, written out explicitly in Appendix~\ref{app:potential}.
\begin{figure}[!t]
\centering 
\includegraphics[width=0.65\textwidth, height=0.2\textwidth]{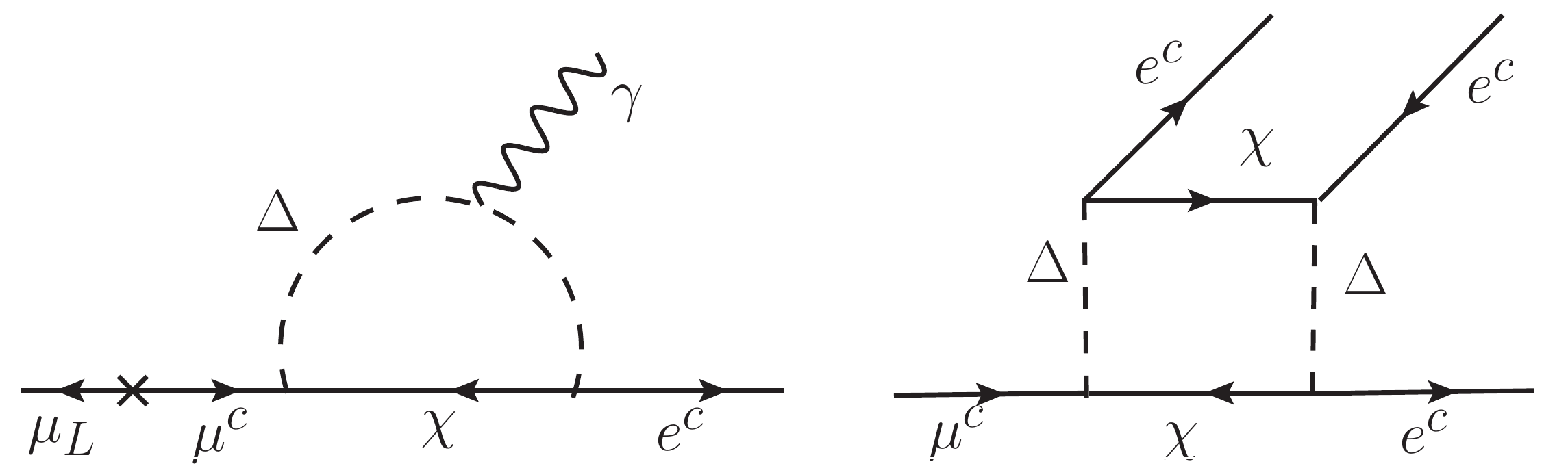}
\caption{Feynman diagrams for the CLFV processes $\mu \to e \gamma$ (left) and $\mu \to 3 \, e$ [box-diagram] (right),  in Model $\chi\Delta\Sigma$.}
\label{fig:NV2_CLFV}
\end{figure}
The operator $\mathcal{O}^{\al\be}_s$ is realized at the tree level with topology 2, and the effective scale is given by 
\begin{equation}
 \frac{g_{\alpha\beta}}{\Lambda^5}\equiv \frac{y_{\Sigma u}\,y_{\Sigma \alpha}\,y_{\Delta \beta}\,y_{\Delta d}^*}{ {M_\Sigma^2\,M_\Delta^2\,m_\chi}}\,.
 \label{Mod2:MutoE}
\end{equation}
The $\mu^-\to e^+$-conversion rates are proportional to $|y_{\Sigma e} y_{\Delta \mu}+y_{\Sigma \mu} y_{\Delta e}|^2$, while those for $0\nu\beta\beta$ are proportional to $|y_{\Sigma e} y_{\Delta e}|^2$.
 
Like the previous example, this model also allows for a rich set of CLFV processes.  The CLFV observables of interest are:
\begin{enumerate}
\item $\mu^{+} \rightarrow e^{+} \gamma$ decay: This is generated at the one-loop level, as depicted in the left panel of Fig.\,\ref{fig:NV2_CLFV}. There is a similar diagram with $\Delta$ and $\chi^c$ in the loop. The effective Lagrangian for this process is
 \begin{equation}
   \mathcal{L}_{\mu^+\to e^+\ga} = \frac{y_{\Delta\mu}\,y_{\Delta e}^*\,(2e)\,y_\mu }{16 \pi^2\,\Lambda^2} (L \overline{H})\,\sigma^{\alpha\beta}e^c F_{\alpha\beta}\,, 
 \end{equation}
 where $\Lambda$ is a function of $M_{\Delta}$ and $m_{\chi}$. The bounds for this model are similar to the ones calculated in Eq.\,(\ref{eq:mutoegamma}).

\item $\mu^{\pm}\rightarrow e^{\pm} e^{\pm} e^{\mp}$ decay: Unlike the previous model, here $\mu\to 3e$ only occurs at the one-loop level. One contribution is obtained from the diagram in the left panel of Fig.\,\ref{fig:NV2_CLFV}, where the photon is off-shell and can ``decay'' into an $e^+e^-$ pair.  As far as this contribution is concerned, the rate for $\mu\to 3e$ is suppressed relative to that for the $\mu\to e\gamma$ decay. There are also box-diagrams, including the one depicted in the right panel of Fig.\,\ref{fig:NV2_CLFV}, which could also contribute significantly. Fig.\,\ref{fig:NV2_CLFV}(right) gives rise to the  effective Lagrangian
\begin{equation}
   \mathcal{L}_{\mu\to 3e} = \frac{y_{\Delta \mu}\,y_{\Delta  e}\,y_{\Delta e}^*\,y_{\Delta e}^*}{16\pi^2 \Lambda^2}\, \left(\overline{\mu^c} e^c\right) \left(\overline{e^c} e^c\right)\,,
\end{equation}
where $\Lambda$ is a function of $M_\Delta$ and $m_\chi$. Using Eq.\,(\ref{eq:mu3e}), current data constrain $\Lambda \geq 23\,{\rm TeV}$ assuming order one couplings.  A similar box-diagram exists with $\chi^c$ and $\Sigma$ in the loop; its contribution turn out to be of the same order. 

\item $\mu^-\to e^-$-conversion in nuclei: In this model, $\mu^-\to e^-$-conversion also occurs at the one-loop level, as depicted in Fig.\,\ref{fig:NV2_mutoe}. The effective Lagrangian can be estimated as
 \begin{equation}
  \mathcal{L}_{\mu\to e} =\left( \frac{y_{\Delta \mu}^*\,y_{\Delta d}\,y_{\Delta d}^*\,y_{\Delta e}}{16\pi^2 \Lambda_{\Delta\chi}^2} +  \frac{y_{\Sigma \mu}^*\,y_{\Sigma u}^*\,y_{\Sigma u}\,y_{\Sigma e}}{16\pi^2 \Lambda_{\Sigma \chi}^2} \right)\, \left(\overline{\mu^c} e^c\right) \left(\overline{d^c} d^c\right)\,
 \end{equation}
where the subscripts on $\Lambda$ denotes the dependence on the masses of the new particles. As in the previous model, this process can also proceed through the transition-magnetic-moment channel, where the photon emits a quark-antiquark pair. The bounds arising on the effective scale for $\mathcal{O}(1)$ couplings are similar to ones obtained in the previous model (Model $\zeta\Phi\Sigma$). There exists a dimension-ten operator $(d^c d^c \overline{d^c}\,\overline{d^c} \ell^c [\sigma\cdot\partial]\overline{\ell^c})$, which can be dressed as the process $n\mu^-\rightarrow n e^- $. The relevant amplitude is 
 \begin{equation}
   \mathcal{A} =\frac{y_{\Delta d}\,y_{\Delta d}^*y_{e\Delta}\,y_{\mu\Delta}^*}{M_{\Delta}^4\,m_\chi^2}(d^c d^c \overline{d^c}\, \overline{d^c} \ell^c [p\cdot\sigma] \overline{\ell^c})
 \end{equation}
 where $p$ is the typical four-momentum associated to the process. Note that there is an analogous contribution to $p\mu^-\to pe^-$. This will also mediate $\mu^-\to e^-$-conversion in nuclei. However, this is an effective operator of very high energy-dimension and hence suppressed. 
 \begin{figure}[t!]
\centering 
\includegraphics[clip,width=0.35\linewidth]{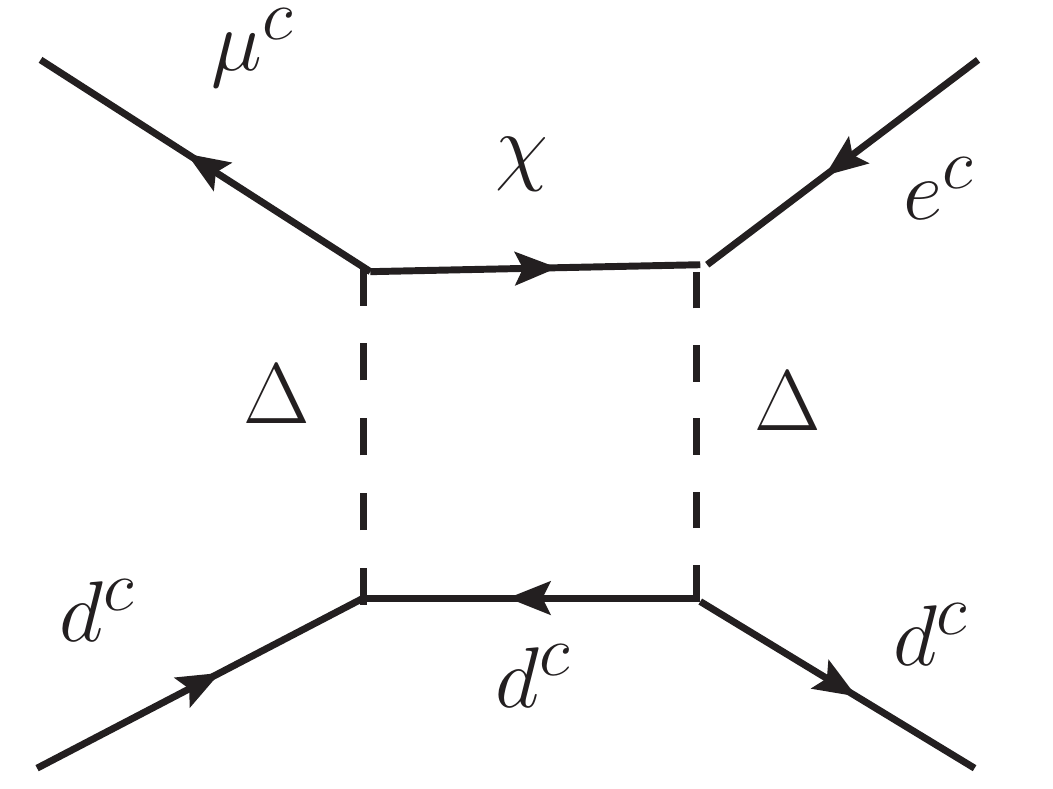}\,\,\,\includegraphics[clip,width=0.35\linewidth]{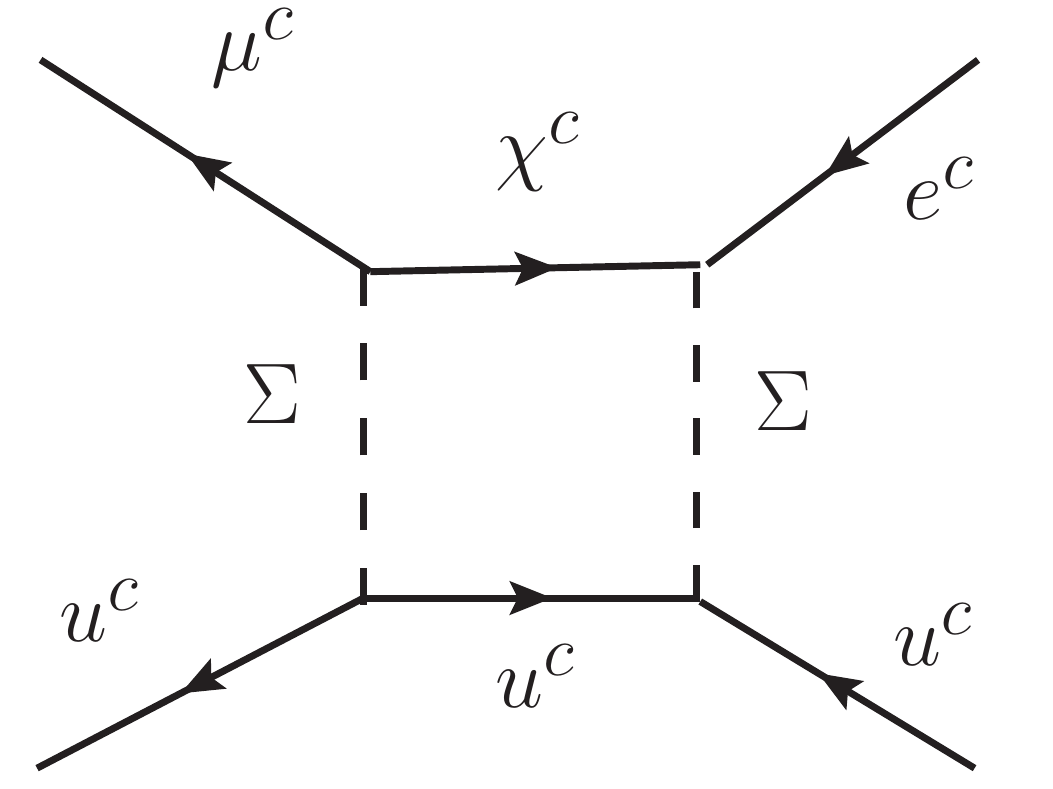}
\caption{Feynman diagrams (box-diagrams) contributing to the CLFV process 
$\mu^- \to e^-$-conversion, in Model $\chi\Delta\Sigma$.}
\label{fig:NV2_mutoe}
\end{figure}

\item Muonium-Antimuonium oscillations and lepton scattering: Unlike the previous model (Model $\zeta\Phi\Sigma$), this model does not allow for tree-level muonium-antimuonium oscillation, or lepton--lepton scattering. One can, of course, have these processes at the one-loop level through diagrams like the right panel of Fig.~\ref{fig:NV2_CLFV}. The bounds arising from these processes are not expected to be competitive with the other leptonic bounds. 

\item Anomalous magnetic moments: there is a new-physics contribution to the anomalous magnetic moment of the muon and the electron at one-loop (e.g., a $\Delta,\chi$ loop). The situation here is very similar to the one discussed in Model $\zeta\Phi\Sigma$. 

\end{enumerate}

A subset of the bounds estimated here are summarized in Fig.~\ref{fig:summaryBounds}. As in the previous model, in the absence of flavor-structure in the new-physics sector, CLFV constraints, along with those from $0\nu\beta\beta$-searches, overwhelm the sensitivity of future searches for $\mu^-\to e^+$-conversion. In this model, it is also possible to consistently assign  $L_\mu - L_e$ charges to the heavy fields and therefore eliminate the processes listed
above. For example, if we assign charge $+1$ to $\chi$ and charge $-1$ to $\chi^c$, only $\mu^c$ couples to $\chi$ and only $e^c$ couples to $\chi^c$. 
This can automatically prevent the above processes from taking place with a sizable rate. Note that this charge assignment will render some of the other new-physics couplings zero, e.g., $y_{\Delta\chi}$ and $y_{\Delta \chi^c}$.

Unlike model $\zeta\Phi\Sigma$, here baryon number is explicitly violated. We note that the Lagrangian Eq.~(\ref{eq:lagrangianModel5}) has an accidental $Z_2$ symmetry under which all lepton-fields, along with $\chi$ and $\chi^c$, are odd. This implies that nucleon decays into leptons are not allowed (e.g., $p \to \pi^0+e^+$ or $n \to \pi^0+\nu$) and, for example, the proton is stable. There are, nonetheless, a few relevant baryon-number-violating (BNV) constraints:

\begin{enumerate}

\item Neutron-antineutron $(n-\overline{n})$ oscillations: at the tree level, the model mediates neutron--antineutron oscillations, which violate baryon number by two units, as depicted in Fig.~\ref{fig:NV2_nnbar}. The effective Lagrangian for such a process is the dimension-nine operator 
 \begin{equation}
   \mathcal{L}_{n-\overline{n}} =\frac{y_{\Sigma u}y_{\Delta d}^2\,m_{\Sigma\Delta}}{M_{\Delta}^4\,M_{\Sigma}^2}(u^c d^c d^c)^2.
   \label{eq:nnbar}
 \end{equation}
Here $m_{\Sigma\Delta}$ is a parameter in the scalar potential, see Appendix~\ref{app:potential}. The Institut Laue-Langevin (ILL) experiment at Grenoble yields the best bounds on free $n-\overline{n}$ oscillations using neutrons from a reactor source \cite{BaldoCeolin:1994jz}. The bounds are typically quoted on the transition matrix element of the effective Hamiltonian, $\delta m = \langle \overline{n}| {H_{\rm eff}} |n\rangle$ and are 
 \begin{equation}
  \tau_{n-\overline{n}}\equiv \frac{1}{|\delta m|} \gtrsim 10^8\,{\rm sec}\,.
 \end{equation}
  
  \begin{figure}[t!]
\centering 
\includegraphics[clip,width=0.4\linewidth]{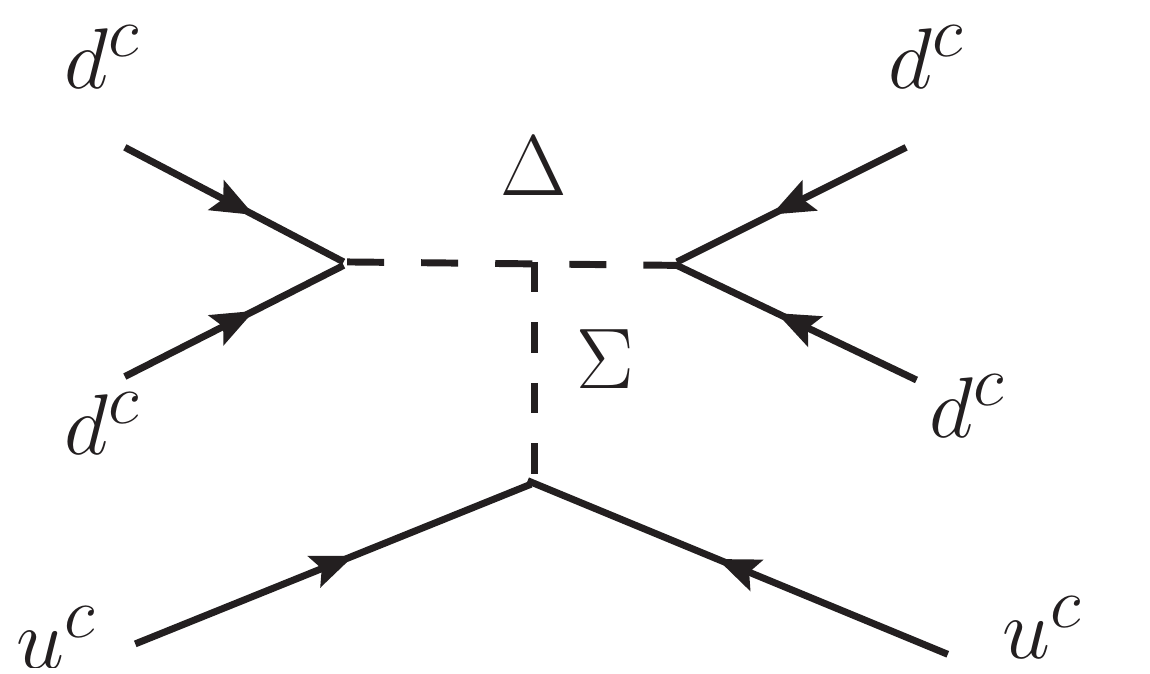}
\caption{Tree-level Feynman diagram that mediates $n-\overline{n}$ oscillations in Model $\chi\Delta\Sigma$.}
\label{fig:NV2_nnbar}
\end{figure}

Using Eq.\,(\ref{eq:nnbar}) \cite{Phillips:2014fgb},
\begin{eqnarray}
\langle \overline{n}| {H_{\rm eff}} |n\rangle&=&\frac{y_{\Sigma u}y_{\Delta d}^2\,m_{\Sigma\Delta}}{M_{\Delta}^4\,M_{\Sigma}^2}\langle \overline{n}|(u^c d^c d^c)^2|n\rangle
       =\frac{y_{\Sigma u}y_{\Delta d}^2\,m_{\Sigma\Delta}}{M_{\Delta}^4\,M_{\Sigma}^2} \Lambda_{\rm QCD}^6\,,
\end{eqnarray}
where we estimate the nucleon matrix-element to be of order $\Lambda_{\rm QCD}$. Assuming $\mathcal{O}(1)$ couplings, $\Lambda_{\rm QCD}=180\,{\rm MeV}$, and $m_{\Sigma\Delta}\sim M_{\Sigma}\sim M_{\Delta}\sim \Lambda$, this translates into 
\begin{equation}
 \Lambda \gtrsim 350\,\,\,{\rm TeV}\,.
\end{equation}

\item BNV processes with LNV: The model also allows for BNV processes that violate lepton number related to the effective dimension-twelve operator $(d^c d^c d^c \overline{\ell^c})^2$ and  
 $(u^c u^c d^c \ell^c)^2$, including $nn\rightarrow \pi^+ \pi^+ e^- e^-$, and $pp\to e^+e^+$. These are expected to be more suppressed given the high energy-dimension of the effective operator. We qualitatively estimate that existing experimental bounds on $pp\to e^+e^+$ \cite{Tanabashi:2018oca} translate into $\Lambda\gtrsim 1$~TeV. 
\end{enumerate}

 The $n-\bar{n}$-oscillation bound also outshines the sensitivity of future $\mu^-\to e^+$-conversion experiments and cannot be avoided by allowing a non-trivial flavor structure to the new-physics since we are especially interested in first-generation quarks. We do note that tree-level BNV processes vanish in the limit $m_{\Sigma\Delta}\to 0$ and hence can be suppressed if $m_{\Sigma\Delta}$ is smaller than the other mass-scales in the theory. The reason is as follows. If we assign baryon number +2/3 to $\Sigma$ and $\Delta$ and $\pm1/3$ to $\chi,\chi^c$ (in units where the quarks have lepton number 1/3), baryon number is violated by the interactions proportional to $y_{\Sigma\alpha}$, $y_{\Delta \alpha}$ -- by one unit -- and $m_{\Sigma\Delta}$ -- by two units. Furthermore, if we assign lepton-number zero to all the new-physics fields, lepton number is violated by $y_{\Sigma\alpha}$, $y_{\Delta \alpha}$ -- by one unit. This means that if $m_{\Sigma\Delta}$ is zero $n-\bar{n}$-oscillation requires one to rely on the interactions proportional to $y_{\Sigma\alpha}$, $y_{\Delta \alpha}$, which also create or destroy leptons. Since there are no leptons in $n-\bar{n}$-oscillation, these interactions contribute to it only at the loop level. In this case, we still expect strong bounds on $ \Lambda \gtrsim 100$~TeV, similar to the one-loop contribution discussed in the next model (Model $\psi\Delta\Phi$). These can be ameliorated by judiciously assuming a subset of new-physics couplings is small. 
 
\subsubsection{Model $\psi\Delta\Phi$}

Here, the SM particle content is augmented by a couple of colored vector-like quarks $\psi \sim (\overline{3},1)_{4/3}$ and $\psi^c \sim (3,1)_{-4/3}$, a colored exotic scalar $\Delta \sim (6,1)_{-2/3}$, and a doubly-charged scalar $\Phi \sim (1,1)_{-2}$. The Lagrangian is given by
\begin{equation}
 \mathcal{L}_{\psi\Delta\Phi} = \mathcal{L}_{\rm SM} +\mathcal{L}_{\rm kin} + y_{\Phi\al \be}\, \Phi \ell^c_\al \ell^c_\be + y_{\Delta d}\,\Delta d^c d^c + y_{\Phi \psi}\,\overline{\Phi} \psi^c u^c  + y_{\Delta \psi}\,\Delta \psi u^c + m_\psi\,\psi\psi^c + V(\Phi,0,\Delta)+ {\rm h.c}\,.
 	\label{eq:lagrangianModel3}
\end{equation}
where $ \mathcal{L}_{\rm SM}$ is the SM Lagrangian, $\mathcal{L}_{\rm kin}$ contains the kinetic-energy terms for the new particles, and $V(\Phi,0,\Delta)$ is the most general scalar potential involving the scalars $\Delta,\Phi$, written out explicitly in Appendix~\ref{app:potential}.

It is easy to check that this model realizes ${\cal O}^{\alpha\beta}_s$ via topology 2 (Fig.~\ref{fig:Topology}(right)) and
\begin{equation}
 \frac{g_{\alpha\beta}}{\Lambda^5}\equiv \frac{y_{\Phi \alpha\beta}\,y_{\Phi\psi}\,y_{\Delta\psi}\,y_{\Delta d}^*}{ M_\Phi^2\,M_\Delta^2\,m_\psi}\,.
\end{equation}
Here, like in Model $\zeta\Phi\Sigma$, $y_{\Phi \alpha\beta}$ controls the lepton-flavor structure of the model. $\mu^-\to e^+$-conversion rates are proportional to $|y_{\Phi e\mu}|^2$, while those for $0\nu\beta\beta$ are proportional to $|y_{\Phi ee}|^2$.

As far as CLFV is concerned, this model is very similar to  Model $\zeta\Phi\Sigma$ since here and there the presence of the doubly-charged scalar $\Phi$ determines most of the lepton-number conserving phenomenology. Similar to Model $\zeta\Phi\Sigma$, the CLFV bounds can be avoided by assuming the new-physics couplings are not generic. If  the new-physics portion of the Lagrangian respects an $L_\mu - L_e$ global symmetry, all CLFV bounds vanish to a very good approximation.

\begin{figure}[t!]
\centering 
\includegraphics[clip,width=0.5\linewidth]{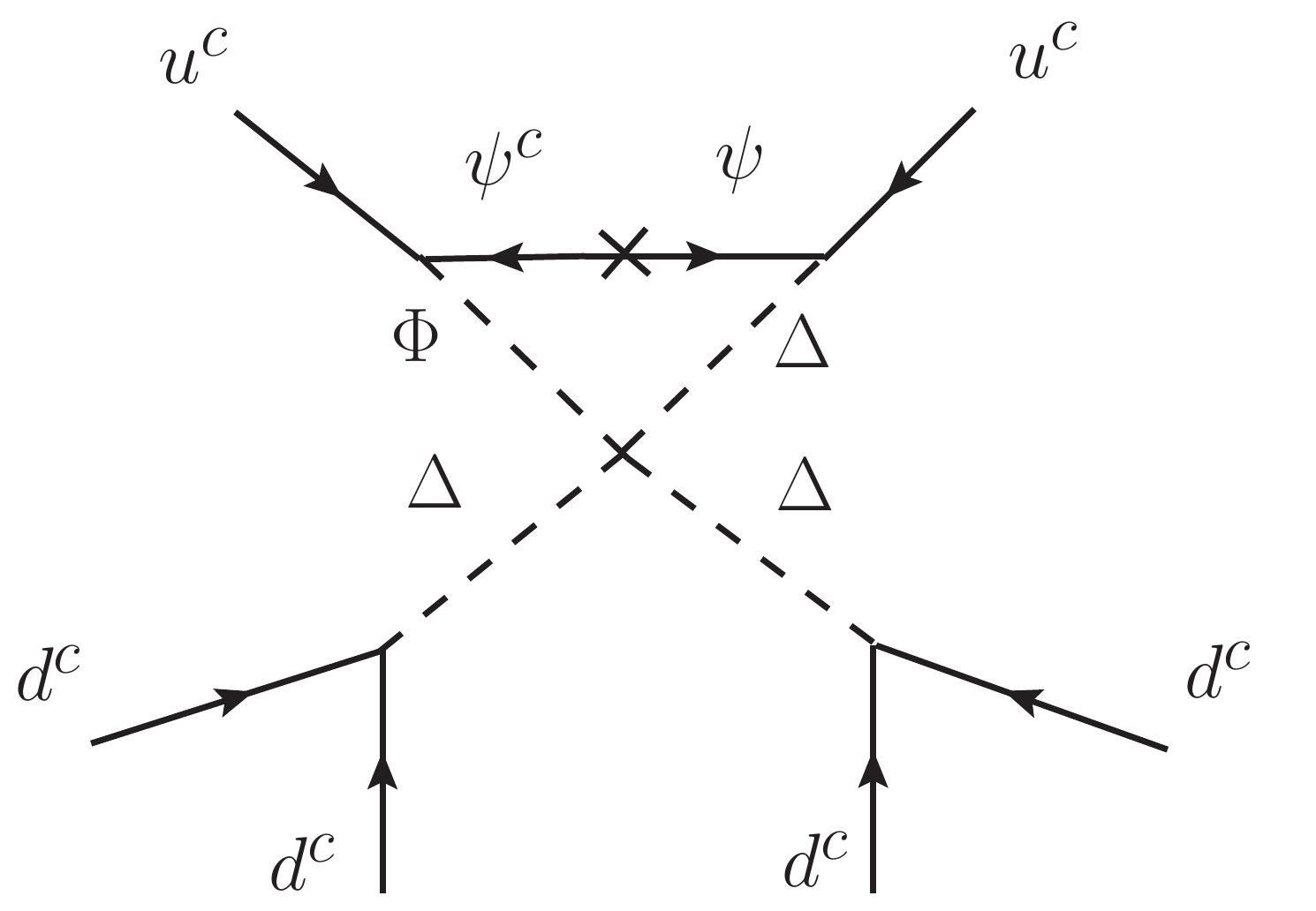}
\caption{One-loop Feynman diagram that mediates $n-\overline{n}$ oscillations in Model $\psi\Delta\Phi$.}
\label{fig:NV3_nnbar}
\end{figure}

Like Model $\chi\Delta\Sigma$, here baryon number is violated but, also like Model $\chi\Delta\Sigma$, there is a $Z_2$ ``lepton-parity'' -- all lepton-fields are odd and all other fields are even -- which implies baryon decays into leptons are not allowed. If we assign lepton-number $+2$ to $\Phi$, baryon-number $+2/3$ to $\Delta$, and baryon-number $\mp1/3$ to $\psi,\psi^c$, baryon-number-violating phenomena are proportional to the $\lambda_{\overline\Delta\Phi}$ coupling in the scalar potential. The same coupling also violates lepton-number by two units.\footnote{From this perspective, the $y_{\Phi\psi}$-coupling violates lepton number by two units.} This implies that $n-\overline{n}$-oscillations do not occur at the tree level since BNV is always accompanied by LNV. However, at one-loop, $n-\overline{n}$-oscillations can take place, as depicted in the Feynman diagram in Fig.\,\ref{fig:NV3_nnbar}. It translates into the effective Lagrangian
\begin{equation}
   \mathcal{L}_{n-\overline{n}} =\frac{y_{\Delta d}^2\,y_{\Phi\psi}\,y_{\Delta\psi}\,\lambda_{\overline{\Delta}\Phi}\,m_{\Psi}}{16\pi^2\,M_{\Delta}^4\,\Lambda^2}(u^c d^c d^c)^2\,,
   \label{eq:nnbarMod3}
 \end{equation}
where $\Lambda$ is an effective scalar arising out of the masses of $\Delta,\,\Phi$ and $\psi$. Assuming all couplings are $\mathcal{O}(1)$ and all mass scales are of the same order, current experimental bounds translate into
\begin{equation}
 \Lambda \gtrsim 127\,\,\,{\rm TeV}\,.
\end{equation}
As advertised, however, baryon-number violation is proportional to $\lambda_{\overline\Delta\Phi}$ and can be suppressed -- or eliminated completely -- in the limit $\lambda_{\overline\Delta\Phi}\to 0$, when baryon number is a good symmetry of the Lagrangian. 

As in Model $\chi\Delta\Sigma$, here one can also construct the dimensional-twelve operator $(d^c d^c d^c \overline{\ell^c})^2$ which gives rise to phenomena like $nn\rightarrow \pi^+ \pi^+ e^- e^-$. Such processes are higher dimensional, and hence expected to be more strongly suppressed. A subset of the bounds, estimated here and in the previous subsubsections, are summarized in Fig.~\ref{fig:summaryBounds}.

\subsubsection{Model $\Phi \Sigma \Delta$}

Here, the SM particle content is augmented by only scalar fields: a color-singlet doubly-charged scalar $\Phi  \sim (1,1)_{-2}$, and two colored scalars, $\Sigma \sim (6,1)_{4/3}$ and $\Delta \sim (6,1)_{-2/3}$. The most general renormalizable Lagrangian is
\begin{equation}
 \mathcal{L}_{ \Phi \Sigma \Delta} = \mathcal{L}_{\rm SM} +\mathcal{L}_{\rm kin} +y_{\Phi\al \be}\,\Phi \ell_\al^c \ell_\be^c + y_{\Delta d}\,\Delta d^c d^c + y_{\Sigma u}\, \Sigma u^c u^c + V(\Phi,\Sigma,\Delta)+ {\rm h.c}\,.
 	\label{eq:lagrangianModel7}
\end{equation}
where $ \mathcal{L}_{\rm SM}$ is the SM Lagrangian, $\mathcal{L}_{\rm kin}$ contains the kinetic-energy terms for the new particles, and $V(\Phi,\Sigma,\Delta)$ is the most general scalar potential involving the scalars $\Phi,\Delta,\Sigma$, written out explicitly in Appendix~\ref{app:potential}.

This is the only no-vectors model where the effective operator $O^{\al\be}_s$ is realized at the tree level through topology 1, and the effective scale is given by
\begin{equation}
 \frac{g_{\alpha\beta}}{\Lambda^5}\equiv \frac{y_{\Phi \alpha\beta}\,y_{\Sigma u}\,y_{\Delta d}^*\,m_{\Delta \Sigma \Phi}^*}{ {M_\Phi^2\,M_\Delta^2\,M_\Sigma^2}}\,.
\label{eq:model4}
\end{equation}
Here, like in Model $\zeta\Phi\Sigma$ and Model $\psi\Delta\Phi$, $y_{\Phi \alpha\beta}$ controls the lepton-flavor structure of the model. $\mu^-\to e^+$-conversion rates are proportional to $|y_{\Phi e\mu}|^2$, while those for $0\nu\beta\beta$ are proportional to $|y_{\Phi ee}|^2$. The CLFV phenomenology here is very similar to the one in Model $\zeta\Phi\Sigma$ and Model $\psi\Delta\Phi$. 

If we choose to assign lepton-number $+2$ to $\Phi$ and baryon-number $+2/3$ to both $\Sigma$ and $\Delta$, all LNV and BNV couplings are in the scalar potential. Some couplings violate only baryon number (e.g., $m_{\Sigma\Delta}$), some violate only lepton number (e.g., $m_{\Delta\Sigma\Phi}$),\footnote{Note that the effective coupling of $O^{\al\be}_s$, Eq.~(\ref{eq:model4}), is proportional to $m_{\Delta\Sigma\Phi}$.} while others violate both (e.g., $\lambda_{\overline{\Delta}\Phi}$). This means that BNV phenomena can occur at the tree level, like in Model $\chi\Delta\Sigma$. Indeed, $n-\overline{n}$-oscillations occur at the tree level via the Feynman diagram in Fig.~\ref{fig:NV2_nnbar}.

\begin{figure}[t!]
\centering 
\includegraphics[clip,width=\linewidth]{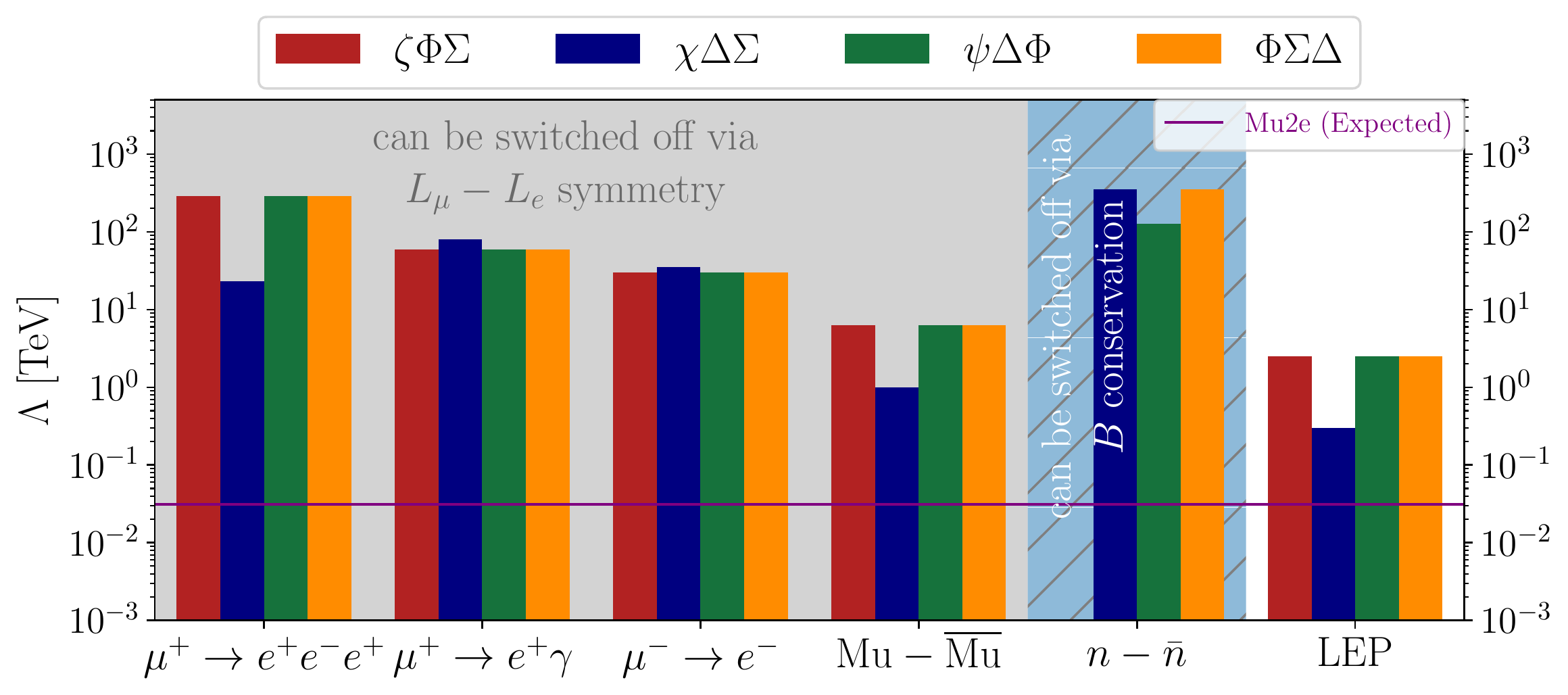}
\caption{Summary of the most stringent CLFV,  BNV, and lepton-scattering bounds on the effective scale of the all-singlets operator for the different models discussed in the text. These bounds assume that all new physics couplings are of order one and all new physics masses are approximately the same. Other bounds are discussed in the text. Bounds from the observables in the grey area can be softened or eliminated if the new-physics couplings have a very non-generic lepton-flavor structure (e.g., if the new-physics model obeys, at least approximately, an $L_{\mu}-L_e$ symmetry, as discussed in the text). Bounds from $n-\bar{n}$-oscillations, in the blue area, can be softened or eliminated if the new-physics couplings are chosen in a way that baryon number is at least approximately conserved. Note that model $\zeta\Phi\Sigma$ conserves baryon number and hence does not contribute to $n-\overline{n}$ oscillations. In the limit where the masses of the new particles are heavy, there are independent hadron collider bounds similar to those from LEP. The expected sensitivity of the {\it Mu2e} experiment is indicated by the solid line.}
\label{fig:summaryBounds}
\end{figure}

A subset of the bounds, estimated in the previous subsubsections, are summarized in Fig.~\ref{fig:summaryBounds}. Here too BNV phenomena are controlled by a different set of couplings as LNV ones, and can be ``turned off'' by imposing baryon number as a conserved, or approximately conserved, symmetry.

\subsection{Models with a New Vector Boson}

As discussed earlier and summarized in Table~\ref{tab:finallist}, there are two different vector bosons capable of realizing the all-singlets operator at the tree level in a way that other LNV operators are also avoided. These are a color-singlet with hyper-charge one [$(1,1)_1$] or a color-octet with hyper-charge one [$(8,1)_1$]. We will refer to both of them as $C_\mu$. The only allowed couplings of $C^{\mu}$ to SM fermions is $C_\mu \overline{d^c} \overline{\sigma}^\mu u^c$ (see Table\,\ref{tab:finallist}). If, however, left-handed antineutrino fields $\nu^c$ exist, the following coupling is also allowed, for the color-singlet $C^{\mu}$: $C_\mu \overline{\ell^c}\overline{\sigma}^\mu\nu^c$. We return to the issue of generating neutrino masses in Sec.~\ref{sec:conclusion}.

Quantum field theories with massive vector bosons, in general, have severe problems in the ultraviolet. The models presented here are no exception. The most general ``UV-complete'' Lagrangians we will be considering are, in fact, not really UV-complete as, for example, we expect the scattering of longitudinal vector bosons to violate partial-wave unitarity in the ultraviolet, indicating that a proper UV-completion of the theory is required. As is well known, there are a few possible ways to UV-complete theories with massive vector bosons. They could, for example, be composite objects of some confining gauge theory. In the scenarios discussed here, since the vector-boson $C^{\mu}$ carries electric-charge (and hyper-charge) and, in some cases, color, some of the fundamental fields of the UV theory must transform nontrivially under the SM gauge symmetry. Another possibitlity is that $C^{\mu}$ is a gauge boson associated to some broken gauge symmetry. The fact that $C^{\mu}$ is charged and potentially colored makes the construction of UV-complete models nontrivial. Below -- in Model $NC$ -- we explore in a little more detail the possibility that the color-singlet $C^{\mu}$ may be the $W_R$-boson in left-right symmetric extensions of the SM. 

All models are listed below. It turns out that, unlike the no-vectors models, all of them conserve baryon number. Phenomenologically, most of the models give rise to the CLFV processes already discussed before and the bounds and challenges one needs to address are very similar to those of no-vectors models. For this reason, we do not elaborate on experimental bounds but, for the most part, concentrate on whatever unique features the different models possess.

\subsubsection{Models $\Phi C$}\label{subsec:model8a}

Here, the SM particle content is augmented by a charged-scalar $\Phi  \sim(1,1)_{-2}$, and a vector $C^\mu\sim (8,1)_1$. The most general renormalizable Lagrangian is 
\begin{equation}
	\mathcal{L}_{\Phi C} =  \mathcal{L}_{\rm SM} + \mathcal{L}_{\rm kin}
	+ y_{\Phi\al \be}\,\Phi \ell_\al^c \ell_\be^c +  g_{Cdu}\,C_\mu \overline{d^c} \overline{\sigma}^\mu u^c + V(\Phi,C) + {\rm h.c.}\,,
	\label{eq:lagrangianModel8a}
\end{equation}
where $ \mathcal{L}_{\rm SM}$ is the SM Lagrangian, $\mathcal{L}_{\rm kin}$ contains the kinetic-energy terms for the new particles, and $V(\Phi,C)$ is the vector-scalar potential listed in Eq.\,(\ref{eq:VPotential}) in Appendix~\ref{app:potential}. This is the simplest model as far as its particle content is concerned.  Lepton number can be assigned to the various fields in a way that the term $C_\mu C^\mu \Phi$ in the vector-scalar potential violates it by two units ($\Phi$ lepton-number 2, $C^{\mu}$ lepton-number zero). 

The all-singlets operator is realized at the tree level via topology 1. The effective couplings and scale are 
\begin{equation}
 \frac{g_{\alpha\beta}}{\Lambda^5}\equiv \frac{y_{\Phi \alpha\beta}\,g_{Cdu}^2\,m_{C\Phi}^*}{ {M_\Phi^2\,M_C^4}}\,.
\end{equation}
Here, like all models that include the $\Phi$-field, $y_{\Phi \alpha\beta}$ controls the lepton-flavor structure of the model. $\mu^-\to e^+$-conversion rates are proportional to $|y_{\Phi e\mu}|^2$, while those for $0\nu\beta\beta$ are proportional to $|y_{\Phi ee}|^2$.

A very similar Lagrangian describes the model where the gauge boson is a color-singlet, $C_\mu \sim (1,1)_1$. The only difference is the presence of an extra interaction between $C^{\mu}$ and the Higgs doublet, proportional to $\overline{C}_\mu H D^\mu H$. This interaction is inconsequential for LNV. 

There are strong constraints on the production of charged vector bosons that couple to quarks, which will be discussed later, while, as already mentioned, the color-singlet vector also allows couplings to left-handed antineutrinos $\propto C_\mu \overline{\ell^c}\overline{\sigma}^\mu\nu^c$.

\subsubsection{Model $\zeta \Phi C$  and $\psi\Phi C$}

We can add a new vetor-like fermion to Model $\Phi C$ in such a way that more LNV interactions are allowed and one generates, at the tree level, the all-singlets operator via both topologies in Fig.~\ref{fig:Topology}. This can be done in two different ways. 

We can add to the particle content of Model $\Phi C$ a pair of vector-like quarks $\zeta \sim (\overline{3},1)_{-5/3}$ and $\zeta^c\sim
(3,1)_{5/3}$. The most general Lagrangian is, assuming $C^{\mu}$ is a color-octet vector-boson,
\begin{align}
	\mathcal{L}_{\zeta\Phi C}=\mathcal{L}_{\rm SM} + \mathcal{L}_{\rm kin}
	+ y_{\Phi\al \be}\,\Phi \ell_\al^c \ell_\be^c  + g_{Cdu} C_\mu \overline{d^c} \overline{\sigma}^\mu u^c + y_{\Phi \zeta^c} \Phi \zeta^c d^c + g_{Cu\zeta} C_\mu \overline{u^c} \overline{\sigma}^\mu \zeta + m_\zeta \zeta\zeta^c + V(\Phi,C)+{\rm h.c.},
	\label{eq:lagrangianModel2a}
\end{align}
where $ \mathcal{L}_{\rm SM}$ is the SM Lagrangian, $\mathcal{L}_{\rm kin}$ contains the kinetic-energy terms for the new particles, and $V(\Phi,C)$ is the vector-scalar potential listed in Eq.\,(\ref{eq:VPotential}) in Appendix~\ref{app:potential}. The coefficient of the all-singlets operator is 
\begin{equation}
 \frac{g_{\alpha\beta}}{\Lambda^5}\equiv \frac{y_{\Phi \alpha\beta}\,g_{Cdu}^2\,m_{C\Phi}^*}{ {M_\Phi^2\,M_C^4}}+ \frac{y_{\Phi \alpha\beta}\,g_{Cdu}\,g_{Cu\zeta}^*\, y_{\Phi \zeta^c}^*}{ {M_\Phi^2\,M_C^2\,m_\zeta}} .
\end{equation}

Instead, we could add to the particle content of Model $\Phi C$ a pair of vector-like quarks  $\psi \sim
(\overline{3},1)_{4/3}$ and $\psi^c\sim (3,1)_{-4/3} $. The most general Lagrangian in this case is, assuming $C^{\mu}$ is a color-octet vector-boson,
\begin{align}
	\mathcal{L}_{\psi\Phi C} = \mathcal{L}_{\rm SM} + \mathcal{L}_{\rm kin}
 + y_{\Phi\al \be}\,\Phi \ell_\al^c \ell_\be^c +  g_{Cdu} C_\mu \overline{d^c} \overline{\sigma}^\mu u^c  + y_{\Phi \psi} \overline{\Phi} \psi^c u^c + g_{C\psi d}\,C_\mu \overline{\psi} \overline{\sigma}^\mu d^c  + m_\psi \psi\psi^c + V(\Phi,C)+ {\rm h.c.},
	\label{eq:lagrangianModel4a}
\end{align}
where $ \mathcal{L}_{\rm SM}$ is the SM Lagrangian, $\mathcal{L}_{\rm kin}$ contains the kinetic-energy terms for the new particles, and $V(\Phi,C)$ is the vector-scalar potential listed in Eq.\,(\ref{eq:VPotential}) in Appendix~\ref{app:potential}. Clearly, this is very similar to the model in Eq.~(\ref{eq:lagrangianModel2a}), with just the charges for the vector-like quarks different. Here, the coefficient of the all-singlets operator is 
\begin{equation}
 \frac{g_{\alpha\beta}}{\Lambda^5}\equiv \frac{y_{\Phi \alpha\beta}\,g_{Cdu}^2\,m_{C\Phi}^*}{ {M_\Phi^2\,M_C^4}}+ \frac{y_{\Phi \alpha\beta}\,g_{Cdu}\,g_{C\psi d}^*\, y_{\Phi \psi}}{ {M_\Phi^2\,M_C^2\,m_\psi}}\,.
\end{equation}

In both scenarios one can assign lepton number to the new-physics fields such that both $m_{C\Phi}$ and the coupling of the $\Phi$ field to the new fermion and a quark -- $y_{\Phi\zeta^c} $ or $y_{\Phi\psi}$ -- violate lepton number by two units. In this way, one can control which topology contributes most to the all-singlets operator. Note, however, that both contributions to $g_{\alpha\beta}/\Lambda^5$ are proportional to $y_{\Phi\alpha\beta}/(M_{\Phi}^2M_C^2)$.

\subsubsection{Model NC}

The new vector-boson $C_\mu \sim (8,1)_1$ can also be used to generate the all-singlets operator at the tree level if there are color-octet fermions $N   \sim (8,1)_0$. In this case, the most general renormalizable Lagrangian is
\begin{equation}
 \mathcal{L}_{NC}= \mathcal{L}_{\rm SM} + \mathcal{L}_{\rm kin} + g_{CN\alpha}\, C_\mu \overline{\ell^c_{\alpha}} \overline{\sigma}^\mu N + g_{Cdu}\,C_\mu \overline{d^c} \overline{\sigma}^\mu u^c + m_N NN + V(0,C)+ {\rm h.c}\,,
	\label{eq:lagrangianModel6a}
\end{equation}
where $ \mathcal{L}_{\rm SM}$ is the SM Lagrangian, $\mathcal{L}_{\rm kin}$ contains the kinetic-energy terms for the new particles, and $V(0,C)$ is the potential for the vector field listed in Eq.\,(\ref{eq:VPotential}) in Appendix~\ref{app:potential}. One can assign lepton number to the new fields, $-1$ for $N$, zero for $C_{\mu}$, such that the Majorana masses of the color-octet fermions control LNV. The operator $O^{\al\be}_s$ is generated at the tree level -- topology 2 -- and its coefficient is
\begin{equation}
 \frac{g_{\alpha\beta}}{\Lambda^5}\equiv \frac{g_{Cdu}^2\,g_{CN\alpha}^{*}g_{CN\beta}^{*}}{ M_C^4 m_N}\,.
\end{equation}
Here, the lepton-flavor structure of the all-singlets operator is governed by the couplings $g_{CN\alpha}$. The $\mu^-\to e^+$-conversion rates are proportional to $|g_{CNe} g_{CN\mu}|^2$, while those for $0\nu\beta\beta$ are proportional to $|g_{CNe}^2|^2$.

Similar to many of the previous models, CLFV process are ubiquitous here. However, since $\mu^c$ and $e^c$ couple to the same fields through the operators $C_\mu\overline{\ell^c} \overline{\sigma}^\mu N$, and since the rate for $\mu^-\to e^+$-conversion requires both $g_{CNe}, g_{CN\mu} $ to be relevant, it is not possible to choose new physics couplings such that most CLFV observables are relatively suppressed. In this scenario, given several existing experimental constraints, the rates for $\mu^-\to e^+$-conversion are outside the reach of the next-generation experiments. However, it is possible to slightly modify the model to suppress CLFV.  Instead of introducing one field $N$, one can introduce the pair $N$ and $N^c$ with the $L_\mu - L_e$ charges $+1$ and $-1$ respectively; in other words, the Lagrangian will include terms $C_\mu \overline{e^c}
\overline{\sigma}^\mu N$ and $C_\mu \overline{\mu^c} \overline{\sigma}^\mu N^c$. The Majorana mass term would be forbidden by the global symmetry and replaced with  the Dirac mass term proportional to $N N^c$.

A similar scenario arises with $C_\mu \sim (1,1)_1$ and a gauge-singlet fermion $N \sim (1,1)_0$. In this case, a neutrino Yukawa interaction $LHN$  is also allowed and the model is nothing more than the type-I seesaw model \cite{Minkowski:1977sc,Yanagida:1979as,Glashow:1979nm,GellMann:1980vs,Mohapatra:1979ia,Schechter:1980gr} plus a charge-one vector boson. This scenario violates the requirements we introduced earlier: here, the Weinberg operator $(LH)^2$ is generated at the tree level, as in the type-I seesaw model. It should be pointed out that it is possible to suppress the tree-level contribution to the Weinberg operator by choosing very small neutrino Yukawa couplings. In this case, the phenomenology is similar to the one discussed in the previous models.\footnote{Here, dimension-seven operators like $L_\mu H \overline{e^c} \overline{u^c}d^c$ are also generated. These yield large contributions to neutrino masses if the Yukawa couplings are not small. In the case of the color-octet Majorana fermion $N \sim (8,1)_0$, Yukawa couplings to the charged leptons do not exist and therefore this is not an issue.}

As discussed before, models with a heavy vector-boson require extra care in order to be rendered consistent in the ultraviolet. In the case of $C_\mu \sim (1,1)_1$, this can be achieved by appreciating that it acts like the right-handed W-boson $W_R$ in left-right symmetric models~\cite{Mohapatra:1974hk,Mohapatra:1974gc,Senjanovic:1975rk,Mohapatra:1979ia,Mohapatra:1980yp}. In fact, the Lagrangian for $C_\mu \sim (1,1)_1$ and the gauge-singlet fermion $N \sim (1,1)_0$ is a subset of the left-right symmetric Lagrangian, where the SM gauge group is extended to $SU(2)_L\times SU(2)_R \times U(1)_{B-L}$. This model requires an extended Higgs sector to break the $ SU(2)_R \times U(1)_{B-L} \rightarrow U(1)_Y$. To avoid bounds from collider experiments, this breaking needs to happen at a higher scale so that the new gauge bosons $W_R,\, Z_R\,$s have large-enough mass. Candidate charge-assignments of the particles under ${\rm SU}(2)_L\times {\rm SU}(2)_R \times {\rm U}(1)_{B-L}$ are listed in Table~\ref{tab:LRSM_par}, where we associate $N$ to the conjugate of the right-handed neutrino $\nu_R$. 

\begin{table}[!h]
\caption{Fields in Model $NC$ assuming $C_{\mu}$ is the right-handed $W_R$-boson of an $SU(2)_L \times SU(2)_R \times U(1)_{B-L}$ gauge theory.}
\begin{center}
 	\begin{tabular}{|p{5cm}|p{5cm}|}
 	\hline
	    Particles  &  $\bigl({\rm SU}(2)_L,\,{\rm SU}(2)_R,\, {\rm U}(1)_{B-L}\bigr)$ \\
	   \hline
	   $Q_L\equiv(u_L,d_L)$ & $ (2,1,1/6)$\\
	   $Q_R\equiv(u_R,d_R)$ & $(1,2,1/6)$\\
	       $\psi_L\equiv(\nu_L,e_L) $ & $(2,1,-1/2)$\\
	       $\psi_R\equiv(\nu_R,e_R) $ & $(1,2,-1/2)$\\
	       $\Delta_L \equiv\text{scalar}  $ & $(3,1,1)$\\
	      $\Delta_R \equiv\text{scalar} $ & $(1,3,1)$\\
	      $\Phi_{LR} \equiv\text{scalar}$ & $(2,2^*,0)$\\
	   \hline
	\end{tabular}
\end{center}
\label{tab:LRSM_par}
\end{table}

The vev of  $\Delta_{R}$, the $SU(2)_{R}$ scalar triplet, gives Majorana masses to the right-handed neutrinos, while that of the of the $SU(2)_L$ scalar triplet $\Delta_L$ contributes to the Majorana masses of the left-handed neutrinos. One can constuct Yukawa interactions involving the Higgs bi-doublet $\Phi_{LR}$, which leads to the $LHN$ Yukawa interaction. In this analogy, $C^\mu \equiv W^{\mu+}_R$, and the interactions $C_\mu \overline{\ell^c} \overline{\sigma}^\mu N, C_\mu \overline{d^c} \overline{\sigma}^\mu u^c$ are gauge interactions.

\setcounter{equation}{0}
\section{Collider Bounds}
\label{sec:coll}

Here we briefly discuss interesting signatures and constraints we expect from collider experiments; a detailed collider study of all models listed in the previous section is beyond the scope of this paper.  All new physics particles introduced in the different models are listed in Table\,\ref{tab:finallist}. They include colored vector-like fermions, charged and colored scalars, and charged and colored vector-bosons. 

As mentioned earlier, the $\Phi$-scalar will mediate $e^+ e^- \to e^+ e^-$ or $e^+ e^- \to \mu^+ \mu^-$ in the $t$-channel. In the limit where the $\Phi$ mass is larger than the center-of-mass energy of the collider, these interactions are already constrained by measurements at LEP~\cite{LEP:2003aa}. For lighter masses, different, stringent constraints on the new-physics couplings are expected. Future $e^+ e^-$ colliders under consideration, like the ILC~\cite{Baer:2013cma}, FCC-ee~\cite{Gomez-Ceballos:2013zzn}) and CEPC~\cite{CEPC-SPPCStudyGroup:2015csa}, would be sensitive to much higher effective mass-scales. The ILC, for example, with an integrated luminosity of 1000 fb$^{-1}$, is capable of probing new physics scales $\Lambda$ that are roughly below 75~TeV~\cite{Riemann:2001bb,Baer:2013cma} (or $M_{\Phi} \lesssim 20$ TeV for order one couplings) through the process $e^+ e^- \to  \mu^+ \mu^-$. The exact sensitivity would depend on the polarization of the electron and positron beams as well as
systematic uncertainties  at the ILC. An $e^-e^-$ collider would be sensitive to $\Phi$ $s$-channel exchange and the properties of $\Phi$ could be studied -- or constrained -- on-resonance if the collider energy were high enough.

The colored scalars $\Sigma$ and $\Delta$, and $C_{\mu}$ (both the color-singlet and the color-octet) can be produced at hadron colliders like the LHC through the quark or the gluon channels. For example, the dijet channel $qq\rightarrow \Sigma\,(\Delta) \rightarrow qq$ can be used to probe the contact interaction $(y^2/M_{\Sigma\,(\Delta)}^2)q^c\,q^c\overline{q^c}\,\overline{q^c}$, which are a valid description of colored-scalar exchange in the limit where the scalar masses are beyond the reach of the collider. Recent dijet studies at ATLAS and CMS \cite{Tanabashi:2018oca,Aaboud:2018fzt,Sirunyan:2018xlo} translate into a lower bound on the mass of scalar diquarks~\cite{Hewett:1988xc} and, in our case, imply masses for $\Sigma$, $\Delta$, and $C_{\mu}$ that exceed around 5 TeV, for order one couplings. $\Sigma$ and $\Delta$ will also mediate, at the tree level,  processes like $gg\rightarrow \Sigma\Sigma\, (\Delta\Delta) \rightarrow 4q$. The corresponding signature is a pair of dijet resonances and can be used to constrain the properties of the colored scalars. The corresponding bounds, however, are expected to be weaker than those of dijet searches as long as the couplings between the new bosons and the quarks are order one. Note that the few TeV upper bound does not trivially apply for smaller couplings and lower masses for $\Sigma$ and $\Delta$ and $C_{\mu}$, see, for example, \cite{Aaboud:2018fzt,Sirunyan:2018xlo}. Relatively-light bosons that couple to quarks relatively strongly are know to survive collider constraints, see for example, \cite{Carone:1994aa}. A detailed analysis of this very rich topic, as mentioned above, is beyond the scope of this paper. 

The literature on searches for vector-like exotic quarks -- including octet ``neutrinos'' --  is also large and diverse. Bounds, many of which are listed and briefly discussed in the particle data book \cite{Tanabashi:2018oca}, hover around 500~GeV. A more detailed discussion of exotic quark searches in the LHC can be found, for example, in \cite{Okada:2012gy}. Existing bounds depend rather strongly on the decay properties of the exotic colored fermions. Model-independent bounds are much weaker, as summarized, for example, in \cite{Tanabashi:2018oca}.

New colored~(and/or charged) particles that couple to the SM Higgs boson will modify the Higgs production rate via gluon fusion and the decay rate into two photons, i.e., $gg \to H \to \ga \ga$. The doubly-charged scalar $\Phi$ also contributes to the decay process $H \to 4\ell$ at tree level.
Precision measurements of Higgs production and decay will translate into bounds
on the properties of $\Phi$, $\Delta$, $\Sigma$, and $C_{\mu}$. In addition, one should also worry about electroweak precision tests of
the SM, although corresponding constraints might be weaker than direct searches
at the LHC. The scalar $\Phi$ and the vector $C_\mu$, for example, will modify (via
triangle loop diagrams) the partial decay widths of the $Z$-boson into quarks
and leptons. Moreover, if $\Phi$ only couples to the pair $e\mu$, the
universality of $Z \to ee$, $\mu\mu$ and $\tau\tau$ will be violated. Finally,
as all new charged particles listed in Table~\ref{tab:finallist} are singlets
under $SU(2)_L$, there are no contributions to the oblique parameters~($S$, $T$,
$U$)~\cite{Peskin:1990zt, Peskin:1991sw}, as demonstrated in, e.g.,
\cite{Lavoura:1992np}, where contributions from vector-like down-type quarks to
the oblique parameters were shown to vanish if these do not mix with the SM
$SU(2)_L$ quark doublets.

LNV phenomena can also be probed at colliders. The all-singlets operator will mediate $\overline{u^c} \overline{u^c} \rightarrow e^c e^c \overline{d^c}\,\overline{d^c}$ scattering, as discussed briefly in \cite{deGouvea:2007qla}. Up to color factors and symmetry factors, the cross section for this process is $\sigma \propto g^2 s^4/\Lambda^{10} $. This can lead to interesting signatures at the LHC or the ILC (exchanging the role of the charged-leptons and the up-quarks). The latter is similar to searches for  the LNV process $e^- e^- \to W^- W^-$ at lepton colliders, except for the fact that the final-state dijet invariant masses are not related to the $W$-boson mass \cite{Kom:2011nc}. Similar studies could also be pursued with a muon collider \cite{Rodejohann:2010jh}.

\section{Summary: Viability of large $\mu^-\to e^+$ conversion rates}
Concerning the viability of different models to mediate observable $\mu^-\to e^+$-conversion in nuclei at next-generation experiments, our main results, illustrated in Fig.~\ref{fig:summaryBounds} with the assumptions of a universal mass scale $\Lambda$ for new particles and $\mathcal{O}(1)$ couplings, can be summarized as follows:
\begin{enumerate}
\item  For all the models considered, CLFV and  $0\nu\beta\beta$  provide the most stringent bounds on the effective scale of the all-singlet operator. These bounds, $\mathcal{O}(10-100)\,{\rm TeV}$, are much stronger than the sensitivity of next-generation $\mu^-\to e^+$-conversion experiments, $\mathcal{O}(10)\,{\rm GeV}$.  However, depending on the lepton-flavor structure of the models considered, it is possible to avoid most of these constaints. One possibility discussed here is that if the new-physics Lagrangian respects an $L_\mu - L_e$ (muon-number minus electron-number) global symmetry, then all the CLFV and  $0\nu\beta\beta$ are significantly weakened. 
\item For models which explicitly violate baryon number, the $n-\bar{n}$-oscillation bound -- $\mathcal{O}(100)\,{\rm TeV}$ -- also outshines the sensitivity of future $\mu^-\to e^+$-conversion experiments and cannot be avoided by allowing a non-trivial flavor structure to the new physics. Even in these cases, one can get remove these bounds by postulating that baryon number is a global symmetry of the Lagrangian.
\item The new interactions predicted by the models are also tightly constrained by LEP and other collider experiments, which also probe scales ($\mathcal{O}(1)\,{\rm TeV}$ ) beyond the sensitivity of future $\mu^-\to e^+$-conversion experiments. These bounds cannot be alleviated by taking advantage of symmetry arguments. They can, however, be weakened by judiciously choosing different couplings (mass-scales) to be relatively small (large), as we discuss in the some concrete scenarios.
\end{enumerate}

\section{Discussions and concluding remarks}
\label{sec:conclusion}

Lepton number and baryon number are accidental global symmetries of the classical SM Lagrangian (and baryon-number--minus-lepton-number is an accidental global symmetry of the quantum SM Lagrangian). LNV can be probed in a variety of ways, ranging from rare nuclear processes to collider experiments. So far, there is no direct evidence for LNV. Nonzero neutrino masses are often interpreted as evidence for LNV. In most scenarios where this is the case, because neutrino masses are tiny, the rates for LNV processes are way out of the reach of experimental probes of LNV, except for searches for $0\nu\beta\beta$. 

Here, we concentrated on identifying and discussing models where this is not the case and asked whether there are UV-complete models where the rate for $\mu^-\to e^+$-conversion in nuclei is close to the sensitivity of next-generation experiments. All models identified here violate lepton number at energies scales around one TeV (or lower) and are best constrained by searches for CLFV, BNV, and $0\nu\beta\beta$. BNV bounds are sometimes strongly correlated, sometimes not, to the LNV physics. LNV scales that are low enough so one approaches the sensitivity of future searches for $\mu^-\to e^+$-conversion in nuclei -- along with other LNV process we did not discuss, like rare meson decays (e.g., $D^-\to K^+\mu^-\mu^-$) -- require a non-generic, but often easy to impose, lepton-flavor structure for the new physics. In these cases, high-energy hadron and lepton colliders also offer interesting constraints and opportunities for future discovery. 

In more detail, we identified all UV-complete models that realize, at low-energies, the all-singlets dimension-nine operator $\mathcal{O}_s=e^c \mu^c u^c u^c \overline{d^c}\, \overline{d^c}$, identified in \cite{Berryman:2016slh}, and do not realize any other LNV effective operator with similar strength. All new particles -- scalars, fermions, and vector bosons -- are listed in Table~\ref{tab:finallist}. Different models consist of the most general renormalizable Lagrangian of the SM plus different combinations of two or three of these particles. Given a concrete Lagrangian, we estimate the rates for and existing constraints from many low-energy observables. The bounds presented here are rough estimates. For the most part, we assume new-physics couplings to be order one, and assume all new mass scales are of the same order.

Given the various bounds estimated here, it is fair to ask whether, for any of the models identified, it is reasonable to assume that the rate for $\mu^-\to e^+$-conversion is within reach of next-generation experiments. The answer, we believe, is affirmative as long as the lepton-flavor structure of the model is not generic and, in some cases, if BNV phenomena are more suppressed than naively anticipated, i.e., BNV couplings are relatively small. At face-value, flavor-independent bounds -- see, for example,  Fig.~\ref{fig:summaryBounds} -- appear to be strong enough to render $\mu^-\to e^+$-conversion out of experimental reach for the foreseeable future. This need not be the case, for a few reasons. One is that the different bounds usually apply only to the masses of a subset of the new-physics particles, while the coefficient of the all-singlets operator depends on the mass of all new degrees-of-freedom. If one saturates all existing bounds carefully, the scale of the all-singlets operator is lower than the strongest lepton-number conserving bounds, depicted in Fig.~\ref{fig:summaryBounds}. Another important point is that, for example, the LEP bounds apply to $y^2/M^2$ in the limit where $M$ is outside the direct reach of LEP. The coefficient of $\mathcal{O}_s$, however, is proportional to $y/M^2$ (see, for example, Eq.~(\ref{Mod1:MutoE}), proportional to $y_{\Phi\mu e}/M_{\Phi}^2$, versus Eq.~(\ref{eq:scatt})), proportional to $y_{\Phi\mu e}^2/M_{\Phi}^2$). For smaller coupling and mass and fixed $y^2/M^2$, $y/M^2$ is relatively larger. Finally, strictly speaking, all estimates here rely on effective theories. For light-enough new particles and smaller couplings, constraints are, in some cases, significantly weaker once translated into the effective scale of the all-singlets operator  $\mathcal{O}_s$.

All of the scenarios discussed here fail, by design, to explain the observed active neutrino masses. CLFV constraints alone imply that the contribution of these new-physics models to Majorana active neutrino masses are tiny, smaller than what is required by observations by at least two or three orders of magnitude. In order to accommodate large active neutrino masses, more degrees-of-freedom, different from the ones discussed here, need to be added to the SM particle content. One possibility is to postulate that, other than the new-physics that leads to the all-singlets operator at the tree level, there are other sources of LNV, perhaps at a much larger energy scale. The high-scale type-I seesaw, with gauge-singlet fermions $\nu^c$ with Majorana masses much larger than the weak scale would do the trick, for example. Most other models constructed to ``explain'' small active neutrino Majorana masses should also work out fine. In some cases, the two sources of LNV may ``interfere,'' as would be the case of the type-I seesaw with any of the models that contain the color-singlet vector boson $C^{\mu}\sim (1,1)_1$. 

Another possibility is to postulate that the physics responsible for the all-singlets operator is the only source of LNV. In this case, small neutrino masses can be accommodated by adding gauge-singlet fermions $\nu^c$ without a Majorana mass and tiny Yukawa couplings to $L$ and $H$. The absence of the Majorana masses for the left-handed antineutrinos is natural in the t'Hooft sense: if the LNV parameters in the models discussed here vanish, lepton number is a good symmetry of the Lagrangian. In this case, neutrinos are pseudo-Dirac fermions since the left-handed neutrinos and the left-handed antineutrinos both acquire small Majorana masses\footnote{In the case of the right-handed neutrinos, their loop-induced Majorana masses are proportional to the neutrino Yukawa couplings. Since the Yukawa couplings are very small, so are the Majorana masses, in spite of the fact that they are associated to a relevant operator.} on top of the dominant Dirac masses. These scenarios are constrained, quite severely, by solar neutrino experiments -- see \cite{deGouvea:2009fp,Donini:2011jh} -- since they mediate neutrino-oscillation processes with long oscillation lengths. A more detailed analysis is beyond the scope of this paper. 

In summary, UV models which induce $\mathcal{O}_s$ at the tree-level can yield a $\mu^- \to e^+$-conversion rate that is accessible to future experiments if
$(i)$ the UV physics respects, at least approximately, a  lepton-flavor symmetry, such as $L_\mu - L_e$, in order to avoid LFV constraints, $(ii)$ the UV  physics respects, at least approximately, baryon-number conservation, in order to evade BNV bounds and $(iii)$ the UV model contains relatively small couplings, especially those that govern lepton-flavor-conserving observables, in order to avoid constraints like those from LEP.

\section*{Acknowledgements}
We would like to thank Jeff Berryman and Kevin Kelly for useful discussions. 
The work of AdG was supported in part by DOE grant \#de-sc0010143. 
WCH acknowledges funding from the Independent Research Fund Denmark, grant number  DFF 6108-00623. 
JK would like to thank Claudia Hagedorn and the Department of Physics, Chemistry and Pharmacy, as well as the Study Travel Fund of the Faculty of Science at the University of Southern Denmark, for their support for his visit at Northwestern University. He would also like to thank Northwestern University and the particle theory group for their hospitality during his stay.
MS acknowledges support from the National Science Foundation, Grant PHY-1630782, and to the Heising-Simons Foundation, Grant 2017-228.
The CP3-Origins centre is partially funded by the Danish National Research Foundation, grant number DNRF90.

\appendix
\section{Scalar and Vector-Scalar Potentials}
\label{app:potential}

The most general potential involving all the Higgs and the new scalars $\Phi  \sim (1,1)_{-2}$, $\Delta \sim (6,1)_{-2/3}$ and $\Sigma \sim (6,1)_{4/3}$, in the no-vectors models, is given by 
\begin{align}
 V(\,\Phi,\,\Sigma,\,\Delta) = & \mu_\Phi^2 |\Phi|^2 + \mu_\Sigma^2  |\Sigma|^2 + \mu_\Delta^2 |\Delta|^2 +\lambda_\Phi |\Phi|^4 +  \lambda_\Sigma |\Sigma|^4 +  \lambda_\Delta |\Delta|^4 + \lambda_{H\Phi}|H|^2 |\Phi|^2 + \lambda_{H\Sigma}|H|^2 |\Sigma|^2 \nonumber\\
               + & \lambda_{H\Delta} |H|^2 |\Delta|^2 + \lambda_{\Phi\Sigma} |\Phi|^2 |\Sigma|^2 + \lambda_{\Phi\Delta}  |\Phi|^2 |\Delta|^2 + \lambda_{\Sigma\Delta}  |\Sigma|^2 |\Delta|^2  + m_{\Sigma \Delta}\,\Sigma \Delta^2 + \lambda_{\overline{\Delta}\Phi}\,\overline{\Delta}^3 \Phi
                \nonumber\\
               + & m_{\Delta\Sigma\Phi}\,\overline{\Delta} \Sigma \Phi + \lambda_{\Delta\Sigma\Phi} \Delta \Sigma^2 \Phi \,. 
 \label{eq:NVPotential}
\end{align}
In the text, we also refer to  $V(\,\Phi,\,\Sigma,\,0)$, $V(\, 0,\,\Sigma,\,\Delta)$, and $V(\,\Phi,\,0,\,\Delta)$. These are given by Eq.~(\ref{eq:NVPotential}) where the field labelled $0$ is set to zero.

Similarly, the most general potential in the vector models involving the Higgs, the scalar $\Phi$ and the vector $C^\mu$ is 
\begin{equation}
  V(\,\Phi,C^\mu) =\mu_\Phi^2 |\Phi|^2 + \mu_C^2 |C_\mu|^2 + \lambda_\Phi |\Phi|^4 + \lambda_C |C_\mu|^4 +  \lambda_{H\Phi} |H|^2 |\Phi|^2 + \lambda_{HC}|H|^2 |C_\mu|^2 + \lambda_{\Phi C} |\Phi|^2 |C_\mu|^2 +  m_{C\Phi}\,C_\mu C^\mu \Phi\,\,.
  \label{eq:VPotential}
\end{equation}
In the text, we also refer to $V(\,0, C^\mu)$. This is given by Eq.~(\ref{eq:VPotential}) where the $\Phi$ field is set to zero.

Throughout, we assume the parameters of the various scalar and scalar-vector potentials are such that none of the new-physics scalar fields acquire vacuum expectation values.

\bibliographystyle{kpmod}
\bibliography{mue.bib}

\begingroup\raggedright\begin{thebibliography}{79}
\expandafter\ifx\csname natexlab\endcsname\relax\def\natexlab#1{#1}\fi

\bibitem['t~Hooft(1976)]{tHooft:1976rip}
G.~'t~Hooft, ``{Symmetry Breaking Through Bell-Jackiw Anomalies}'', {\em Phys.
  Rev. Lett.} {\bfseries 37} (1976) 8--11,
[,226(1976)].

\bibitem[Klinkhamer and Manton(1984)]{Klinkhamer:1984di}
F.~R. Klinkhamer and N.~S. Manton, ``{A Saddle Point Solution in the
  Weinberg-Salam Theory}'', {\em Phys. Rev.} {\bfseries D30} (1984)
2212.

\bibitem[Ellis and Sakurai(2016)]{Ellis:2016ast}
J.~Ellis and K.~Sakurai, ``{Search for Sphalerons in Proton-Proton
  Collisions}'', {\em JHEP} {\bfseries 04} (2016) 086,
 \href{https://arxiv.org/abs/1601.03654}{{\ttfamily arXiv:1601.03654}}.

\bibitem[Tye and Wong(2015)]{Tye:2015tva}
S.~H.~H. Tye and S.~S.~C. Wong, ``{Bloch Wave Function for the Periodic
  Sphaleron Potential and Unsuppressed Baryon and Lepton Number Violating
  Processes}'', {\em Phys. Rev.} {\bfseries D92} (2015), no.~4, 045005,
 \href{https://arxiv.org/abs/1505.03690}{{\ttfamily arXiv:1505.03690}}.

\bibitem[Tanabashi et~al.(2018)]{Tanabashi:2018oca}
{\bfseries Particle Data Group} Collaboration, M.~Tanabashi {\em et~al.},
  ``{Review of Particle Physics}'', {\em Phys. Rev.} {\bfseries D98} (2018),
  no.~3,
030001.

\bibitem[Schechter and Valle(1982)]{Schechter:1981cv}
J.~Schechter and J.~W.~F. Valle, ``{Neutrino Decay and Spontaneous Violation of
  Lepton Number}'', {\em Phys. Rev.} {\bfseries D25} (1982)
774.

\bibitem[de~Gouv\^ea and Vogel(2013)]{deGouvea:2013zba}
A.~de~Gouv\^ea and P.~Vogel, ``{Lepton Flavor and Number Conservation, and
  Physics Beyond the Standard Model}'', {\em Prog. Part. Nucl. Phys.}
  {\bfseries 71} (2013) 75--92,
 \href{https://arxiv.org/abs/1303.4097}{{\ttfamily arXiv:1303.4097}}.

\bibitem[de~Gouv\^ea(2016)]{Gouvea:2016shl}
A.~de~Gouv\^ea, ``{Neutrino Mass Models}'', {\em Ann. Rev. Nucl. Part. Sci.}
  {\bfseries 66} (2016)
197--217.

\bibitem[Schechter and Valle(1982)]{Schechter:1981bd}
J.~Schechter and J.~W.~F. Valle, ``{Neutrinoless Double beta Decay in SU(2) x
  U(1) Theories}'', {\em Phys. Rev.} {\bfseries D25} (1982) 2951,
[,289(1981)].

\bibitem[Rodejohann(2011)]{Rodejohann:2011mu}
W.~Rodejohann, ``{Neutrino-less Double Beta Decay and Particle Physics}'', {\em
  Int. J. Mod. Phys.} {\bfseries E20} (2011) 1833--1930,
 \href{https://arxiv.org/abs/1106.1334}{{\ttfamily arXiv:1106.1334}}.

\bibitem[Albert et~al.(2014)]{Albert:2014awa}
{\bfseries EXO-200} Collaboration, J.~B. Albert {\em et~al.}, ``{Search for
  Majorana neutrinos with the first two years of EXO-200 data}'', {\em Nature}
  {\bfseries 510} (2014) 229--234,
 \href{https://arxiv.org/abs/1402.6956}{{\ttfamily arXiv:1402.6956}}.

\bibitem[Gando et~al.(2013)]{Gando:2012zm}
{\bfseries KamLAND-Zen} Collaboration, A.~Gando {\em et~al.}, ``{Limit on
  Neutrinoless $\beta\beta$ Decay of $^{136}$Xe from the First Phase of
  KamLAND-Zen and Comparison with the Positive Claim in $^{76}$Ge}'', {\em
  Phys. Rev. Lett.} {\bfseries 110} (2013), no.~6, 062502,
 \href{https://arxiv.org/abs/1211.3863}{{\ttfamily arXiv:1211.3863}}.

\bibitem[Agostini et~al.(2018)]{GERDA:2018zzh}
{\bfseries GERDA} Collaboration, M.~Agostini {\em et~al.}, ``{GERDA results and
  the future perspectives for the neutrinoless double beta decay search using
  $^{76}$Ge}'', {\em Int. J. Mod. Phys.} {\bfseries A33} (2018), no.~09,
1843004.

\bibitem[Bolton and Deppisch(2019)]{Bolton:2019wta}
P.~D. Bolton and F.~F. Deppisch, ``{Probing nonstandard lepton number violating
  interactions in neutrino oscillations}'', {\em Phys. Rev.} {\bfseries D99}
  (2019), no.~11, 115011,
 \href{https://arxiv.org/abs/1903.06557}{{\ttfamily arXiv:1903.06557}}.

\bibitem[Kuno(2013)]{Kuno:2013mha}
{\bfseries COMET} Collaboration, Y.~Kuno, ``{A search for muon-to-electron
  conversion at J-PARC: The COMET experiment}'', {\em PTEP} {\bfseries 2013}
  (2013)
022C01.

\bibitem[Natori(2014)]{Natori:2014yba}
{\bfseries DeeMe} Collaboration, H.~Natori, ``{DeeMe experiment - an
  experimental search for a mu-e conversion reaction at J-PARC MLF}'', {\em
  Nucl. Phys. Proc. Suppl.} {\bfseries 248-250} (2014)
52--57.

\bibitem[Bartoszek et~al.(2014)]{Bartoszek:2014mya}
{\bfseries Mu2e} Collaboration, L.~Bartoszek {\em et~al.}, ``{Mu2e Technical
  Design Report}'',
 \href{https://arxiv.org/abs/1501.05241}{{\ttfamily arXiv:1501.05241}}.

\bibitem[Kaulard et~al.(1998)]{Kaulard:1998rb}
{\bfseries SINDRUM II} Collaboration, J.~Kaulard {\em et~al.}, ``{Improved
  limit on the branching ratio of $\mu^-\to e^+$ conversion on titanium}'',
  {\em Phys. Lett.} {\bfseries B422} (1998)
334--338.

\bibitem[Berryman et~al.(2017)Berryman, de~Gouvêa, Kelly, and
  Kobach]{Berryman:2016slh}
J.~M. Berryman, A.~de~Gouvêa, K.~J. Kelly, and A.~Kobach,
  ``{Lepton-number-violating searches for muon to positron conversion}'', {\em
  Phys. Rev.} {\bfseries D95} (2017), no.~11, 115010,
 \href{https://arxiv.org/abs/1611.00032}{{\ttfamily arXiv:1611.00032}}.

\bibitem[Yeo et~al.(2017)Yeo, Kuno, Lee, and Zuber]{Yeo:2017fej}
B.~Yeo, Y.~Kuno, M.~Lee, and K.~Zuber, ``{Future experimental improvement for
  the search of lepton-number-violating processes in the $e\mu$ sector}'', {\em
  Phys. Rev.} {\bfseries D96} (2017), no.~7, 075027,
 \href{https://arxiv.org/abs/1705.07464}{{\ttfamily arXiv:1705.07464}}.

\bibitem[Geib et~al.(2017)Geib, Merle, and Zuber]{Geib:2016atx}
T.~Geib, A.~Merle, and K.~Zuber, ``{$\mu^- - e^+$ conversion in upcoming LFV
  experiments}'', {\em Phys. Lett.} {\bfseries B764} (2017) 157--162,
 \href{https://arxiv.org/abs/1609.09088}{{\ttfamily arXiv:1609.09088}}.

\bibitem[Geib and Merle(2016)]{Geib:2016daa}
T.~Geib and A.~Merle, ``{$\mu^-$- $e^+$ Conversion from Short-Range
  Operators}'',
 \href{https://arxiv.org/abs/1612.00452}{{\ttfamily arXiv:1612.00452}}.

\bibitem[Babu and Leung(2001)]{Babu:2001ex}
K.~S. Babu and C.~N. Leung, ``{Classification of effective neutrino mass
  operators}'', {\em Nucl. Phys.} {\bfseries B619} (2001) 667--689,
 \href{https://arxiv.org/abs/hep-ph/0106054}{{\ttfamily arXiv:hep-ph/0106054}}.

\bibitem[de~Gouv\^ea and Jenkins(2008)]{deGouvea:2007qla}
A.~de~Gouv\^ea and J.~Jenkins, ``{A Survey of Lepton Number Violation Via
  Effective Operators}'', {\em Phys. Rev.} {\bfseries D77} (2008) 013008,
 \href{https://arxiv.org/abs/0708.1344}{{\ttfamily arXiv:0708.1344}}.

\bibitem[Angel et~al.(2013)Angel, Rodd, and Volkas]{Angel:2012ug}
P.~W. Angel, N.~L. Rodd, and R.~R. Volkas, ``{Origin of neutrino masses at the
  LHC: $\Delta L = 2$ effective operators and their ultraviolet completions}'',
  {\em Phys. Rev.} {\bfseries D87} (2013), no.~7, 073007,
 \href{https://arxiv.org/abs/1212.6111}{{\ttfamily arXiv:1212.6111}}.

\bibitem[Deppisch et~al.(2018)Deppisch, Graf, Harz, and
  Huang]{Deppisch:2017ecm}
F.~F. Deppisch, L.~Graf, J.~Harz, and W.-C. Huang, ``{Neutrinoless Double Beta
  Decay and the Baryon Asymmetry of the Universe}'', {\em Phys. Rev.}
  {\bfseries D98} (2018), no.~5, 055029,
 \href{https://arxiv.org/abs/1711.10432}{{\ttfamily arXiv:1711.10432}}.

\bibitem[Weinberg(1979)]{Weinberg:1979sa}
S.~Weinberg, ``{Baryon and Lepton Nonconserving Processes}'', {\em Phys. Rev.
  Lett.} {\bfseries 43} (1979)
1566--1570.

\bibitem[Esteban et~al.(2019)Esteban, Gonzalez-Garcia, Hernandez-Cabezudo,
  Maltoni, and Schwetz]{Esteban:2018azc}
I.~Esteban, M.~C. Gonzalez-Garcia, A.~Hernandez-Cabezudo, M.~Maltoni, and
  T.~Schwetz, ``{Global analysis of three-flavour neutrino oscillations:
  synergies and tensions in the determination of $\theta_23, \delta_CP$, and
  the mass ordering}'', {\em JHEP} {\bfseries 01} (2019) 106,
 \href{https://arxiv.org/abs/1811.05487}{{\ttfamily arXiv:1811.05487}}.

\bibitem[Vagnozzi et~al.(2017)Vagnozzi, Giusarma, Mena, Freese, Gerbino, Ho,
  and Lattanzi]{Vagnozzi:2017ovm}
S.~Vagnozzi, E.~Giusarma, O.~Mena, K.~Freese, M.~Gerbino, S.~Ho, and
  M.~Lattanzi, ``{Unveiling $\nu$ secrets with cosmological data: neutrino
  masses and mass hierarchy}'', {\em Phys. Rev.} {\bfseries D96} (2017),
  no.~12, 123503,
 \href{https://arxiv.org/abs/1701.08172}{{\ttfamily arXiv:1701.08172}}.

\bibitem[Aghanim et~al.(2018)]{Aghanim:2018eyx}
{\bfseries Planck} Collaboration, N.~Aghanim {\em et~al.}, ``{Planck 2018
  results. VI. Cosmological parameters}'',
 \href{https://arxiv.org/abs/1807.06209}{{\ttfamily arXiv:1807.06209}}.

\bibitem[Loureiro et~al.(2018)]{Loureiro:2018pdz}
A.~Loureiro {\em et~al.}, ``{On The Upper Bound of Neutrino Masses from
  Combined Cosmological Observations and Particle Physics Experiments}'',
 \href{https://arxiv.org/abs/1811.02578}{{\ttfamily arXiv:1811.02578}}.

\bibitem[Gando et~al.(2016)]{KamLAND-Zen:2016pfg}
{\bfseries KamLAND-Zen} Collaboration, A.~Gando {\em et~al.}, ``{Search for
  Majorana Neutrinos near the Inverted Mass Hierarchy Region with
  KamLAND-Zen}'', {\em Phys. Rev. Lett.} {\bfseries 117} (2016), no.~8, 082503,
   \href{https://arxiv.org/abs/1605.02889}{{\ttfamily arXiv:1605.02889}},
[Addendum: Phys. Rev. Lett. {\bf 117}, 109903 (2016)].

\bibitem[Bertl et~al.(2006)]{Bertl:2006up}
{\bfseries SINDRUM II} Collaboration, W.~H. Bertl {\em et~al.}, ``{A Search for
  muon to electron conversion in muonic gold}'', {\em Eur. Phys. J.} {\bfseries
  C47} (2006)
337--346.

\bibitem[Helo et~al.(2015)Helo, Hirsch, Ota, and Pereira~dos
  Santos]{Helo:2015fba}
J.~C. Helo, M.~Hirsch, T.~Ota, and F.~A. Pereira~dos Santos, ``{Double beta
  decay and neutrino mass models}'', {\em JHEP} {\bfseries 05} (2015) 092,
 \href{https://arxiv.org/abs/1502.05188}{{\ttfamily arXiv:1502.05188}}.

\bibitem[Anamiati et~al.(2018)Anamiati, Castillo-Felisola, Fonseca, Helo, and
  Hirsch]{Anamiati:2018cuq}
G.~Anamiati, O.~Castillo-Felisola, R.~M. Fonseca, J.~C. Helo, and M.~Hirsch,
  ``{High-dimensional neutrino masses}'', {\em JHEP} {\bfseries 12} (2018) 066,
 \href{https://arxiv.org/abs/1806.07264}{{\ttfamily arXiv:1806.07264}}.

\bibitem[Bellgardt et~al.(1988)]{Bellgardt:1987du}
{\bfseries SINDRUM} Collaboration, U.~Bellgardt {\em et~al.}, ``{Search for the
  Decay mu+ ---> e+ e+ e-}'', {\em Nucl. Phys.} {\bfseries B299} (1988)
1--6.

\bibitem[Kuno and Okada(2001)]{Kuno:1999jp}
Y.~Kuno and Y.~Okada, ``{Muon decay and physics beyond the standard model}'',
  {\em Rev. Mod. Phys.} {\bfseries 73} (2001) 151--202,
 \href{https://arxiv.org/abs/hep-ph/9909265}{{\ttfamily arXiv:hep-ph/9909265}}.

\bibitem[Fael and Greub(2017)]{Fael:2016yle}
M.~Fael and C.~Greub, ``{Next-to-leading order prediction for the decay $ \mu
  \to e\kern0.22em \left({e}^{+}{e}^{-}\right)\;\nu \kern0.2em \overline{\nu}
  $}'', {\em JHEP} {\bfseries 01} (2017) 084,
 \href{https://arxiv.org/abs/1611.03726}{{\ttfamily arXiv:1611.03726}}.

\bibitem[Berger(2014)]{Berger:2014vba}
{\bfseries Mu3e} Collaboration, N.~Berger, ``{The Mu3e Experiment}'', {\em
  Nucl. Phys. Proc. Suppl.} {\bfseries 248-250} (2014)
35--40.

\bibitem[Baldini et~al.(2016)]{TheMEG:2016wtm}
{\bfseries MEG} Collaboration, A.~M. Baldini {\em et~al.}, ``{Search for the
  lepton flavour violating decay $\mu ^+ \rightarrow \mathrm {e}^+ \gamma $
  with the full dataset of the MEG experiment}'', {\em Eur. Phys. J.}
  {\bfseries C76} (2016), no.~8, 434,
 \href{https://arxiv.org/abs/1605.05081}{{\ttfamily arXiv:1605.05081}}.

\bibitem[Raidal and Santamaria(1998)]{Raidal:1997hq}
M.~Raidal and A.~Santamaria, ``{Muon electron conversion in nuclei versus mu
  $\to$ e gamma: An Effective field theory point of view}'', {\em Phys. Lett.}
  {\bfseries B421} (1998) 250--258,
 \href{https://arxiv.org/abs/hep-ph/9710389}{{\ttfamily arXiv:hep-ph/9710389}}.

\bibitem[Baldini et~al.(2013)]{Baldini:2013ke}
A.~M. Baldini {\em et~al.}, ``{MEG Upgrade Proposal}'',
 \href{https://arxiv.org/abs/1301.7225}{{\ttfamily arXiv:1301.7225}}.

\bibitem[Willmann et~al.(1999)Willmann, Schmidt, Wirtz, Abela, Baranov,
  Bagaturia, Bertl, Engfer, Gro\ss{}mann, Hughes, Jungmann, Karpuchin, Kisel,
  Korenchenko, Korenchenko, Kravchuk, Kuchinsky, Leuschner, Meyer, Merkel,
  Moiseenko, Mzavia, zu~Putlitz, Reichart, Reinhard, Renker, Sakhelashvilli,
  Tr\"ager, and Walter]{PhysRevLett.82.49}
L.~Willmann, P.~V. Schmidt, H.~P. Wirtz, R.~Abela, V.~Baranov, J.~Bagaturia,
  W.~Bertl, R.~Engfer, A.~Gro\ss{}mann, V.~W. Hughes, K.~Jungmann,
  V.~Karpuchin, I.~Kisel, A.~Korenchenko, S.~Korenchenko, N.~Kravchuk,
  N.~Kuchinsky, A.~Leuschner, V.~Meyer, J.~Merkel, A.~Moiseenko, D.~Mzavia,
  G.~zu~Putlitz, W.~Reichart, I.~Reinhard, D.~Renker, T.~Sakhelashvilli,
  K.~Tr\"ager, and H.~K. Walter, ``New bounds from a search for muonium to
  antimuonium conversion'', {\em Phys. Rev. Lett.} {\bfseries 82} Jan (1999)
  49--52.

\bibitem[Electroweak(2003)]{LEP:2003aa}
{\bfseries LEP, ALEPH, DELPHI, L3, OPAL, LEP Electroweak Working Group, SLD
  Electroweak Group, SLD Heavy Flavor Group} Collaboration, t.~S. Electroweak,
  ``{A Combination of preliminary electroweak measurements and constraints on
  the standard model}'',
 \href{https://arxiv.org/abs/hep-ex/0312023}{{\ttfamily arXiv:hep-ex/0312023}}.

\bibitem[Bennett et~al.(2006)]{Bennett:2006fi}
{\bfseries Muon g-2} Collaboration, G.~W. Bennett {\em et~al.}, ``{Final Report
  of the Muon E821 Anomalous Magnetic Moment Measurement at BNL}'', {\em Phys.
  Rev.} {\bfseries D73} (2006) 072003,
 \href{https://arxiv.org/abs/hep-ex/0602035}{{\ttfamily arXiv:hep-ex/0602035}}.

\bibitem[Davier et~al.(2011)Davier, Hoecker, Malaescu, and
  Zhang]{Davier:2010nc}
M.~Davier, A.~Hoecker, B.~Malaescu, and Z.~Zhang, ``{Reevaluation of the
  Hadronic Contributions to the Muon g-2 and to alpha(MZ)}'', {\em Eur. Phys.
  J.} {\bfseries C71} (2011) 1515,
  \href{https://arxiv.org/abs/1010.4180}{{\ttfamily arXiv:1010.4180}},
[Erratum: Eur. Phys. J.C72,1874(2012)].

\bibitem[Hagiwara et~al.(2011)Hagiwara, Liao, Martin, Nomura, and
  Teubner]{Hagiwara:2011af}
K.~Hagiwara, R.~Liao, A.~D. Martin, D.~Nomura, and T.~Teubner, ``{$(g-2)_\mu$
  and $\alpha(M_Z^2)$ re-evaluated using new precise data}'', {\em J. Phys.}
  {\bfseries G38} (2011) 085003,
 \href{https://arxiv.org/abs/1105.3149}{{\ttfamily arXiv:1105.3149}}.

\bibitem[Chakrabarty et~al.(2018)Chakrabarty, Chiang, Ohata, and
  Tsumura]{Chakrabarty:2018qtt}
N.~Chakrabarty, C.-W. Chiang, T.~Ohata, and K.~Tsumura, ``{Charged scalars
  confronting neutrino mass and muon $g-2$ anomaly}'', {\em JHEP} {\bfseries
  12} (2018) 104,
 \href{https://arxiv.org/abs/1807.08167}{{\ttfamily arXiv:1807.08167}}.

\bibitem[Moore et~al.(1985)Moore, Whisnant, and Young]{Moore:1984eg}
S.~R. Moore, K.~Whisnant, and B.-L. Young, ``{Second Order Corrections to the
  Muon Anomalous Magnetic Moment in Alternative Electroweak Models}'', {\em
  Phys. Rev.} {\bfseries D31} (1985)
105.

\bibitem[Lindner et~al.(2016)Lindner, Platscher, and Queiroz]{Lindner:2016bgg}
M.~Lindner, M.~Platscher, and F.~S. Queiroz, ``{A Call for New Physics : The
  Muon Anomalous Magnetic Moment and Lepton Flavor Violation}'',
 \href{https://arxiv.org/abs/1610.06587}{{\ttfamily arXiv:1610.06587}}.

\bibitem[Grange et~al.(2015)]{Grange:2015fou}
{\bfseries Muon g-2} Collaboration, J.~Grange {\em et~al.}, ``{Muon (g-2)
  Technical Design Report}'',
 \href{https://arxiv.org/abs/1501.06858}{{\ttfamily arXiv:1501.06858}}.

\bibitem[Baldo-Ceolin et~al.(1994)]{BaldoCeolin:1994jz}
M.~Baldo-Ceolin {\em et~al.}, ``{A New experimental limit on neutron -
  anti-neutron oscillations}'', {\em Z. Phys.} {\bfseries C63} (1994)
409--416.

\bibitem[Phillips et~al.(2016)]{Phillips:2014fgb}
D.~G. Phillips, II {\em et~al.}, ``{Neutron-Antineutron Oscillations:
  Theoretical Status and Experimental Prospects}'', {\em Phys. Rept.}
  {\bfseries 612} (2016) 1--45,
 \href{https://arxiv.org/abs/1410.1100}{{\ttfamily arXiv:1410.1100}}.

\bibitem[Minkowski(1977)]{Minkowski:1977sc}
P.~Minkowski, ``{$\mu \to e\gamma$ at a Rate of One Out of $10^{9}$ Muon
  Decays?}'', {\em Phys. Lett.} {\bfseries 67B} (1977)
421--428.

\bibitem[Yanagida(1979)]{Yanagida:1979as}
T.~Yanagida, ``{Horizontal gauge symmetry and masses of neutrinos}'', {\em
  Conf. Proc.} {\bfseries C7902131} (1979)
95--99.

\bibitem[Glashow(1980)]{Glashow:1979nm}
S.~L. Glashow, ``{The Future of Elementary Particle Physics}'', {\em NATO Sci.
  Ser. B} {\bfseries 61} (1980)
687.

\bibitem[Gell-Mann et~al.(1979)Gell-Mann, Ramond, and Slansky]{GellMann:1980vs}
M.~Gell-Mann, P.~Ramond, and R.~Slansky, ``{Complex Spinors and Unified
  Theories}'', {\em Conf. Proc.} {\bfseries C790927} (1979) 315--321,
 \href{https://arxiv.org/abs/1306.4669}{{\ttfamily arXiv:1306.4669}}.

\bibitem[Mohapatra and Senjanovic(1980)]{Mohapatra:1979ia}
R.~N. Mohapatra and G.~Senjanovic, ``{Neutrino Mass and Spontaneous Parity
  Nonconservation}'', {\em Phys. Rev. Lett.} {\bfseries 44} (1980) 912,
[,231(1979)].

\bibitem[Schechter and Valle(1980)]{Schechter:1980gr}
J.~Schechter and J.~W.~F. Valle, ``{Neutrino Masses in SU(2) x U(1)
  Theories}'', {\em Phys. Rev.} {\bfseries D22} (1980)
2227.

\bibitem[Mohapatra and Pati(1975{\natexlab{a}})]{Mohapatra:1974hk}
R.~N. Mohapatra and J.~C. Pati, ``{Left-Right Gauge Symmetry and an
  Isoconjugate Model of CP Violation}'', {\em Phys. Rev.} {\bfseries D11}
  (1975){\natexlab{a}}
566--571.

\bibitem[Mohapatra and Pati(1975{\natexlab{b}})]{Mohapatra:1974gc}
R.~N. Mohapatra and J.~C. Pati, ``{A Natural Left-Right Symmetry}'', {\em Phys.
  Rev.} {\bfseries D11} (1975){\natexlab{b}}
2558.

\bibitem[Senjanovic and Mohapatra(1975)]{Senjanovic:1975rk}
G.~Senjanovic and R.~N. Mohapatra, ``{Exact Left-Right Symmetry and Spontaneous
  Violation of Parity}'', {\em Phys. Rev.} {\bfseries D12} (1975)
1502.

\bibitem[Mohapatra and Senjanovic(1981)]{Mohapatra:1980yp}
R.~N. Mohapatra and G.~Senjanovic, ``{Neutrino Masses and Mixings in Gauge
  Models with Spontaneous Parity Violation}'', {\em Phys. Rev.} {\bfseries D23}
  (1981)
165.

\bibitem[Baer et~al.(2013)Baer, Barklow, Fujii, Gao, Hoang, Kanemura, List,
  Logan, Nomerotski, Perelstein, et~al.]{Baer:2013cma}
H.~Baer, T.~Barklow, K.~Fujii, Y.~Gao, A.~Hoang, S.~Kanemura, J.~List, H.~E.
  Logan, A.~Nomerotski, M.~Perelstein, {\em et~al.}, ``{The International
  Linear Collider Technical Design Report - Volume 2: Physics}'',
 \href{https://arxiv.org/abs/1306.6352}{{\ttfamily arXiv:1306.6352}}.

\bibitem[Bicer et~al.(2014)]{Gomez-Ceballos:2013zzn}
{\bfseries TLEP Design Study Working Group} Collaboration, M.~Bicer {\em
  et~al.}, ``{First Look at the Physics Case of TLEP}'', {\em JHEP} {\bfseries
  01} (2014) 164,
 \href{https://arxiv.org/abs/1308.6176}{{\ttfamily arXiv:1308.6176}}.

\bibitem[Ahmad et~al.(2015)]{CEPC-SPPCStudyGroup:2015csa}
M.~Ahmad {\em et~al.}, ``{CEPC-SPPC Preliminary Conceptual Design Report. 1.
  Physics and Detector}'',
2015.

\bibitem[Riemann(2001)]{Riemann:2001bb}
S.~Riemann, ``{Fermion pair production at a linear collider: A Sensitive tool
  for new physics searches}'',
2001.

\bibitem[Aaboud et~al.(2018)]{Aaboud:2018fzt}
{\bfseries ATLAS} Collaboration, M.~Aaboud {\em et~al.}, ``{Search for low-mass
  dijet resonances using trigger-level jets with the ATLAS detector in $pp$
  collisions at $\sqrt{s}=13$ TeV}'', {\em Phys. Rev. Lett.} {\bfseries 121}
  (2018), no.~8, 081801,
 \href{https://arxiv.org/abs/1804.03496}{{\ttfamily arXiv:1804.03496}}.

\bibitem[Sirunyan et~al.(2018)]{Sirunyan:2018xlo}
{\bfseries CMS} Collaboration, A.~M. Sirunyan {\em et~al.}, ``{Search for
  narrow and broad dijet resonances in proton-proton collisions at $
  \sqrt{s}=13 $ TeV and constraints on dark matter mediators and other new
  particles}'', {\em JHEP} {\bfseries 08} (2018) 130,
 \href{https://arxiv.org/abs/1806.00843}{{\ttfamily arXiv:1806.00843}}.

\bibitem[Hewett and Rizzo(1989)]{Hewett:1988xc}
J.~L. Hewett and T.~G. Rizzo, ``{Low-Energy Phenomenology of Superstring
  Inspired E(6) Models}'', {\em Phys. Rept.} {\bfseries 183} (1989)
193.

\bibitem[Carone and Murayama(1995)]{Carone:1994aa}
C.~D. Carone and H.~Murayama, ``{Possible light U(1) gauge boson coupled to
  baryon number}'', {\em Phys. Rev. Lett.} {\bfseries 74} (1995) 3122--3125,
 \href{https://arxiv.org/abs/hep-ph/9411256}{{\ttfamily arXiv:hep-ph/9411256}}.

\bibitem[Okada and Panizzi(2013)]{Okada:2012gy}
Y.~Okada and L.~Panizzi, ``{LHC signatures of vector-like quarks}'', {\em Adv.
  High Energy Phys.} {\bfseries 2013} (2013) 364936,
 \href{https://arxiv.org/abs/1207.5607}{{\ttfamily arXiv:1207.5607}}.

\bibitem[Peskin and Takeuchi(1990)]{Peskin:1990zt}
M.~E. Peskin and T.~Takeuchi, ``{A New constraint on a strongly interacting
  Higgs sector}'', {\em Phys. Rev. Lett.} {\bfseries 65} (1990)
964--967.

\bibitem[Peskin and Takeuchi(1992)]{Peskin:1991sw}
M.~E. Peskin and T.~Takeuchi, ``{Estimation of oblique electroweak
  corrections}'', {\em Phys. Rev.} {\bfseries D46} (1992)
381--409.

\bibitem[Lavoura and Silva(1993)]{Lavoura:1992np}
L.~Lavoura and J.~P. Silva, ``{The Oblique corrections from vector - like
  singlet and doublet quarks}'', {\em Phys. Rev.} {\bfseries D47} (1993)
2046--2057.

\bibitem[Kom and Rodejohann(2012)]{Kom:2011nc}
C.~H. Kom and W.~Rodejohann, ``{Four-jet final state in same-sign lepton
  colliders and neutrinoless double beta decay mechanisms}'', {\em Phys. Rev.}
  {\bfseries D85} (2012) 015013,
 \href{https://arxiv.org/abs/1110.3220}{{\ttfamily arXiv:1110.3220}}.

\bibitem[Rodejohann(2010)]{Rodejohann:2010jh}
W.~Rodejohann, ``{Inverse Neutrino-less Double Beta Decay Revisited: Neutrinos,
  Higgs Triplets and a Muon Collider}'', {\em Phys. Rev.} {\bfseries D81}
  (2010) 114001,
 \href{https://arxiv.org/abs/1005.2854}{{\ttfamily arXiv:1005.2854}}.

\bibitem[de~Gouv\^ea et~al.(2009)de~Gouv\^ea, Huang, and
  Jenkins]{deGouvea:2009fp}
A.~de~Gouv\^ea, W.-C. Huang, and J.~Jenkins, ``{Pseudo-Dirac Neutrinos in the
  New Standard Model}'', {\em Phys. Rev.} {\bfseries D80} (2009) 073007,
 \href{https://arxiv.org/abs/0906.1611}{{\ttfamily arXiv:0906.1611}}.

\bibitem[Donini et~al.(2011)Donini, Hernandez, Lopez-Pavon, and
  Maltoni]{Donini:2011jh}
A.~Donini, P.~Hernandez, J.~Lopez-Pavon, and M.~Maltoni, ``{Minimal models with
  light sterile neutrinos}'', {\em JHEP} {\bfseries 07} (2011) 105,
 \href{https://arxiv.org/abs/1106.0064}{{\ttfamily arXiv:1106.0064}}.

\end{thebibliography}\endgroup

\end{document}